\documentclass[aps,pre,preprint,superscriptaddress]{revtex4-1}

\usepackage[utf8x]{inputenc}
\usepackage{graphicx}
\usepackage[margin=10pt, font=small]{caption}
\usepackage[USenglish]{babel}
\usepackage{amsmath}
\usepackage{amssymb}

\begin{document}

\title{Measuring cellular traction forces on non-planar substrates}
\author{J\'er\^ome R Soin\'e}
\affiliation{Institute for Theoretical Physics, University of Heidelberg, Heidelberg, Germany}
\affiliation{BioQuant, University of Heidelberg, Heidelberg, Germany}
\author{Nils Hersch}
\affiliation{Institute of Complex Systems 7: Biomechanics, Forschungszentrum J\"ulich GmbH, 52425 J\"ulich, Germany}
\author{Georg Dreissen}
\affiliation{Institute of Complex Systems 7: Biomechanics, Forschungszentrum J\"ulich GmbH, 52425 J\"ulich, Germany}
\author{Nico Hampe}
\affiliation{Institute of Complex Systems 7: Biomechanics, Forschungszentrum J\"ulich GmbH, 52425 J\"ulich, Germany}
\author{Bernd Hoffmann}
\affiliation{Institute of Complex Systems 7: Biomechanics, Forschungszentrum J\"ulich GmbH, 52425 J\"ulich, Germany}
\author{Rudolf Merkel}
\affiliation{Institute of Complex Systems 7: Biomechanics, Forschungszentrum J\"ulich GmbH, 52425 J\"ulich, Germany}
\author{Ulrich S. Schwarz}
\affiliation{Institute for Theoretical Physics, University of Heidelberg, Heidelberg, Germany}
\affiliation{BioQuant, University of Heidelberg, Heidelberg, Germany}

\begin{abstract}
Animal cells use traction forces to sense the mechanics and geometry of their
environment. Measuring these traction forces requires a workflow
combining cell experiments, image processing and force reconstruction
based on elasticity theory. Such procedures have been established before
mainly for planar substrates, in which case one can use the Green's function
formalism. Here we introduce a worksflow to measure
traction forces of cardiac myofibroblasts on non-planar elastic substrates.
Soft elastic substrates with a wave-like topology were micromolded from 
polydimethylsiloxane (PDMS) and fluorescent
marker beads were distributed homogeneously in the substrate.
Using feature vector based tracking of these marker beads, we first constructed
a hexahedral mesh for the substrate. We then solved the direct
elastic boundary volume problem on this mesh using the finite element method (FEM).
Using data simulations, we show that the traction forces can be reconstructed
from the substrate deformations by solving the corresponding inverse
problem with a L1-norm for the residue and a L2-norm for 0th order Tikhonov regularization.
Applying this procedure to the experimental data, we find that 
cardiac myofibroblast cells tend to align both their shapes and their
forces with the long axis of the deformable wavy substrate.
\end{abstract}

\maketitle

\section{Introduction}

Animal tissue cells sense the mechanical and geometrical features of their environment by applying traction forces
to the extracellular matrix. Various studies over the last decades demonstrated the importance of  tissue mechanics
for cell behavior, including cell adhesion, migration, proliferation, differentiation and fate \cite{Vogel2006,Chen2008,Humphrey2014}.
It has been shown early that cells respond sensitively to the rigidity of their environment, with larger spreading
area and higher force generation on stiffer substrates \cite{Pelham1997,Engler2004,Nisenholz2014}. Even
more dramatically, essential cell processes such as cell differentiation are controlled by substrate stiffness \cite{McBeath2004,Engler2006,Wen2014}.
Because environmental stiffness is a passive property of the environment, cells have to actively pull on it to determine its
magnitude \cite{schwarz_physics_2013}.

Cellular traction forces are not only used to sense extracellular stiffness, but also
to sense geometrical properties of the environment \cite{Bischofs2005,Kollmannsberger2011,kraning2012controlling}. This is most evident
for adhesive patterns, whose geometry also determines cell responses such as
cell survival and differentiation \cite{Chen1997a,McBeath2004,Kilian2010}.
However, geometry sensing also includes cell sensitivity to topographical features of the environment, which spans
several orders of magnitude, ranging from the molecular up to the cellular scale \cite{Curtis1997,Bettinger2009,Kim2012}.
It has been shown that cell differentiation can be controled also by nano-topography \cite{dalby2007control,yim2007synthetic}.
Regarding the cytoskeletal response, it has been found that cells tend to align with grooves on substrates with corresponding
nano- and micro-topography \cite{dalby2003nucleus,Biela2009,Ulbrich2011}. Another physiological relevant example is
topography-driven polarity guidance and directional growth of neurons \cite{Corey2003}.
Early examples for topography sensing on the \textit{$\mu m$} scale are studies that showed the alignment of actin microfilaments in cells adhered to microcylinders \cite{Svitkina1995,Levina1996}. Interestingly it was found that cells tend to align either parallel or orthogonal to the direction of highest curvature dependent on the cell type. Inspired by these observations mechanical cell models have been proposed that suggest the importance of mechanical stress in the cytoskeleton for the detection and the response to curved structures \cite{Biton2009,Sanz-Herrera2009}.

Recently the combined effect of stiffness and geometry has been studied by engineering topographic features
into polyacrylamide substrates, which as hydrogels however tend to swell in medium and therefore change dimensions \cite{charest2012fabrication}. 
Interestingly, it was found that cells align with the topographic features independent of stiffness.
However, no attempt has been made to measure cellular traction forces in these experiments.
In general, there are many physiological situations in which cells are exposed
to mechanical and geometrical cues simultaneously, and therefore will use cellular
traction forces to sense them both at the same time. An interesting example are podocytes, which are epithelial cells lining the
basement membrane of the glomerular capillaries in kidneys, which have a wavy shape \cite{Endlich2006}.
However, a setup that allows to quantitatively measure forces for such situations is still missing,
despite recent progress in measuring cellular traction forces on planar substrates.

During the last three decades, the measurement of cellular forces (traction force microscopy, TFM) has become a mature research field \cite{Cesa2007,Style2014,Plotnikov2014,Schwarz2015}. The standard setup uses a planar elastic substrate whose deformations are tracked
using embedded marker beads. Standard choices for substrate material are
polyacrylamide (PAA) or crosslinked polydimethylsiloxane (PDMS). For sufficiently thick substrates, one then can use the Green's function of 
an elastic halfspace to relate these deformations to traction forces (GF-TFM) \cite{Dembo1999}. For
a thin substrate, the corresponding Green's function is also known \cite{Merkel2007,DelAlamo2007,Style2014}. TFM-procedures
can be made quite fast by inverting the elastic equations in Fourier space (Fourier transform traction cytometry, FTTC),
which is a special case of GF-TFM \cite{Butler2002}. Traditionally, cell forces
had been reconstructed only in the x-y-plane of a planar substrate (with the z-direction denoting
the normal direction), but over the last years, the Green's function approach has
been extended to also reconstruct the z-forces that cells exert to the substrate \cite{Delanoe-Ayari2010}.  
Alternatively, one can use the finite element method (FEM) to reconstruct these z-forces (FEM-TFM) \cite{Hur2012,Legant2013}.
Within a FEM-approach, one does not rely on the analytical form of a Green's function, but uses numerical 
solutions to the mechanical problem interpolated on a suitable chosen grid \cite{Yang2006,Legant2010,steinwachs2015}.
Another alternative to GF-TFM is the direct method, in which deformations are directly converted
into a stress tensor, from which the traction forces are extracted \cite{Maskarinec2009,Toyjanova2014}.

If one aims at implementing TFM for non-planar substrates, from these three methods (GF-TFM, FEM-TFM and direct TFM)
only the second one seems feasible. First it is notoriously hard to analytically calculate
Green's functions for non-planar (e.g. wavy) surfaces, thus ruling out GF-TFM. As we will see
below, wavy substrates lead to rather noisy displacement data, which is hard to deal with in
direct TFM, because it relies on constructing derivatives of the measured data. Therefore we opted for
an approach using FEM-TFM, which needs more computer time than traditional GF-TFM,
but offers the same level of robustness. We first developed a novel experimental technique
to prepare curved micromolded PDMS substrates with embedded fluorescent marker beads matching the requirements of TFM application.
In contrast to the PAA-hydrogels often used for studies of mechanosensing, the micromolded PDMS-substrates used
here do not suffer from swelling and therefore present time-independent geometrical cues to the cells.
However, because the point spread function in such a substrate varies in space, special
procedures have to be used for tracking the marker beads.
Because here we focus on measuring traction forces, we did not vary the mechanical stiffness of the substrates.
On the computational side we established a complete workflow to reconstruct cellular traction fields on such non-planar substrates. The core of this technique is a parallelized optimization framework that efficiently implements FEM to reconstruct cellular traction forces in 3D. We validated our procedures by reconstructing simulated traction patterns under various experimental conditions. 
In particular, we show that the use of the L1-norm for the definition of the residue strongly improves our force reconstruction
because it better deals with outliers than the L2-norm.
We then applied TFM to cardiac myofibroblasts cultured on curved elastic substrates, thus complementing a traditional contact guidance experiment with measurements of cell traction forces. By comparing both polarization of the cytoskeleton and the distribution of cellular traction we show that cells not only adjust their morphology, but also their moments of traction force to geometrical properties of their surrounding.
 
\section{Methods}

\subsection{Experimental procedures}

Elastomeric substrates were made of Sylgard $184$ Silicone Elastomer Kit (Dow Corning GmbH, Wiesbaden, Germany) as described previously \cite{Cesa2007}. In brief, both components (base and cross-linker) of the elastomer kit were mixed at various ratios to generate after cross-linking either stamps from vinyl disc masters (10:1) or cell culture substrates of stiffness 15 kPa (50:1). For stamp manufacturing 
round polypropylene rings (diameter 10 mm) were placed on top of the vinyl disc and crosslinked at $40^{\circ}$C  for 4 h. Stamps were
peeled off and silanized with tricholorosilane. For force analysis cell culture substrates were equipped with red fluorescent beads (0.1 µm diameter, non-modified beads, Magspheres, CA, USA) resulting in a dense microstructured volume (typical bead
distance 4 $\mu m$). Base oil was mixed with a 1:50 dilution of beads in methanol. The modified base oil was incubated at $60^{\circ}$C for 3 h to evaporate methanol. The desired amount of cross-linker was added to the modified base oil and mixed intensively. The stamp was placed onto a 80 $\mu m$ coverslip with a drop of silicone oil mixture. Layer thickness was adjusted to 80 $\mu m$ using glass slices of same defined thickness (Menzel GmbH, Braunschweig, Germany) as spacers. Cross-linking of silicone oil was performed at $60^{\circ}$C for 16 h. After curing stamps were carefully peeled off and structured elastomeric substrates were glued to the bottom of 3.5 cm Petri-dishes to cover predrilled holes \cite{ulbricht2013cellular}. Mechanical properties of all elastomeric mixing ratios were characterized as described resulting in elasticities as given above and a Poisson's ratio of $0.5$.

Cardiac fibroblasts were isolated from $19$-day-old Wistar rat embryos as described previously \cite{hersch2013constant,Hampe2014}. Primary cardiac fibroblasts were cultured for additional $5$ days at $37^{\circ}$C and $5\%$ CO$_2$ in a humidified incubator on standard polystyrene cell culture dishes to induce their differentiation to myofibroblasts, which have more prominent focal adhesions and stress fibers. Cells were trypsinized and subsequently transfected using Nucleofector technology (Lonza-Amaxa Systems, Cologne, Germany). $10^6$ cells were resuspended in $100~\mu l$ liposome solution containing $2~\mu g$ purified plasmid-DNA (GFP-VASP or GFP-Vinculin). After transfection, resuspended myofibroblasts were seeded on 
fibronectin coated ($2.5$ $~\mu g/cm^2$) (BD Bioscience, New Jersey, USA)
silicone rubber wave substrates at a density of $2$x$10­­^4$ per sample for traction force microscopy. 
The cells were seeded for at least 24 hours before the measurements were taken.

Live cell analyses were performed at $37^{\circ}$C and $5\%$ CO$_2$ using an inverse confocal laser scanning microscope
(LSM 710, Zeiss, Germany, software ZEN 2011) equipped with live cell imaging accessories and a
$63$x Planapochromat oil immersion objective (Ph3, NA $1.4$, Zeiss). Confocal 3D micrographs were taken
using appropriate laser and filter settings for detection of green and red fluorescent light.
Z-stacks of approximately $50~\mu m$ with optimized overlap were taken. As reference value,
z-stacks without cells (peeled off with glass micro needle) were performed with the same parameters.
Acquisition of a single slice took 8 seconds and we typically acquired 100 slices with
a distance around 0.4 $\mu m$. Thus acquisition time for one stack was around 15 minutes.

\subsection{Bead tracking}

Bead tracking for wavy substrates is challenging because of the long stack acquisition time leading to considerable drift
not only between the deformed and the reference states, but even within one stack. We figured that this drift
did not result from cell activity, because 24 hours after spreading, they were quiescent. Moreover the
observed drift was similar at any depth in the substrate. As another special feature of wavy substrates,
we observed that in contrast to planar substrates, the point spread function varied in space.

Our bead tracking routine consists of three steps: determination of image drift, single bead localization, and feature vector based tracking of bead movement. To reduce noise, the stacks were first binomially filtered with a $3$x$3$x$3$ filter kernel, voxel sizes $0.1 \mu$m or $0.11\ \mu$m lateral and $0.39\ \mu$m or $0.56\ \mu$m vertical. Then spots (usually $4$) in the undeformed corners of the image were marked (in the x-y-plane). At these spots a cuboid (VOI, volume of interest) with $10\%$ of the image size in each dimension was cropped from the first image and cross-correlated with the second image to obtain the drift. Because the drift in each slice of the stack can differ, this VOI is cropped along each slice and so the drift is calculated for every slice separately. The area in which the VOI is cross-correlated with the second stack has to be larger than the maximum shift of the two image stacks. To decrease the calculation time for the cross correlation (CC), the size of the image stack and the VOIs are initially reduced by two levels of Gaussian pyramid. The positions of the maximum CC values are then used to calculate the drift on the full image to get the exact drift. This can be done in a very small search area ($5$x$5$x$5$ pixel) around the previously calculated positions. To obtain a subpixel accurate positioning of the VOIs, parabolas were fitted through three points (maximum of CC and both neighbors) in x, y, and z-directions, respectively. The extreme values of the fits are defined as the subpixel accurate VOI positions. The CC and parabola fitting is also done in the first image stack to get a subpixel accurate position there as well. The mean difference between the VOI positions in the first and in the second stack defines the drift, separately for each slice. 

To localize beads in each image stack, both stacks were first filtered with a $7$x$7$ and a $31$x$31$ binomial filter. These filtered images were then subtracted, to get a bandpass-filtered stack, with negative gray values set to zero. Then a VOI that represents one bead in three dimensions is selected manually and fitted to a 3D Gaussian. The fit is used as a reference template to find other beads that match the template. 
The fitted Gaussian is cross-correlated with the first and the second stack. Then the cross-correlated dataset is segmented with a threshold ($0.7$) and each segment is labeled. At the maximum CC value in each labeled segment the data is again fitted 
to a Gaussian and the center of mass of that fit is defined as bead position.

Once the beads are localized in both stacks, they have to be correlated to each other. Here, we adapted a feature vector based tracking method
as published previously \cite{Legant2010}. Before we apply the tracking procedure, the determined drift is subtracted from the bead positions found in the second stack. Then the distance of a bead (B1) from stack 1 to all other beads in its neighborhood ($50$x$50$ pixel in x-y- and $20$ pixel in z- direction) is calculated. To the at most $7$ beads with the lowest distance, vectors from B1 to those beads are created. These vectors are added to beads (B2) in stack 2 (again only in a neighborhood of $50$x$50$x$20$ pixel around the position of B1) and a cubic surrounding (e.g. $7$x$7$x$7$ pixel) is created at each end of the vector. If there is a bead located within this surrounding, a hit is counted. The more hits are counted, the bigger is the probability that bead B1 corresponds to B2. The bead B2 with the most hits (at least $3$ out of $7$) is assigned to bead B1. If two or more beads B2 have the same amount of hits, the one with the lowest deviation at the vector ends is chosen.

\subsection{Reconstruction of substrate shape}

As pre-processing step for the traction reconstruction we need to determine the substrate shapes. Although the substrate preparation described above leads to reproducible samples with $\mu$m accuracy, shape variations and material relaxation after mold removal occasionally change the final shape. The latter effect can in principle be predicted theoretically \cite{Gordan2008}, however, in practice local shape variations occur and make the use of theoretical shape predictions difficult. For that reason we determine the substrate shape for each individual TFM data set by image processing of the relaxed configuration of the marker beads, which are distributed sufficiently homogeneously in the substrate as to carve out the substrate shape. 

\begin{figure}[t]
\centering
\includegraphics[width=\textwidth]{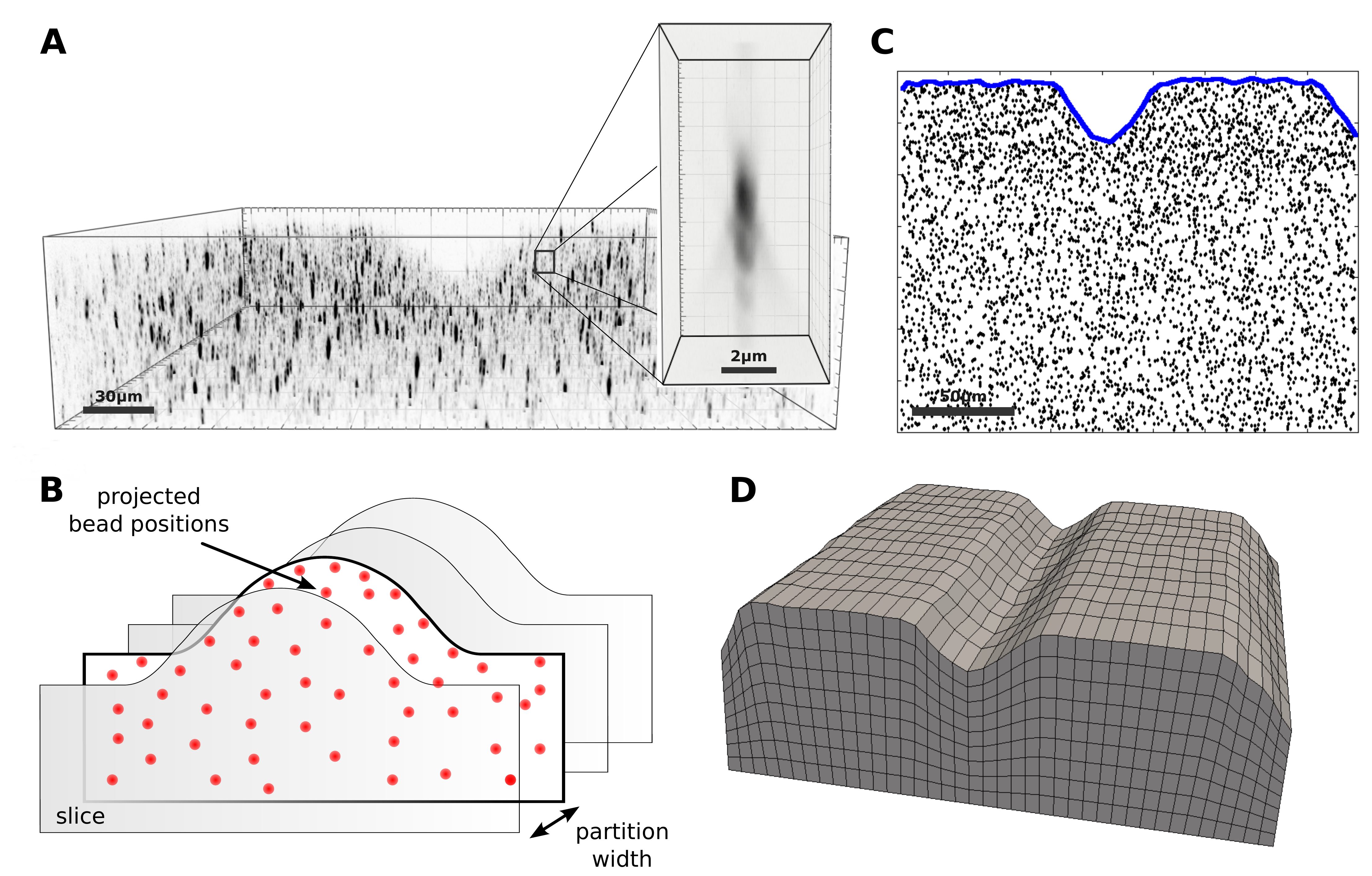}
\caption{Reconstruction of substrate shape from bead positions. 
(A) Image of the marker beads. The inset shows the anisotopic nature of the point spread function.
(B) The substrate contour is partitioned into volumeric slices along a lateral axis. Bead displacements within the partitioned volumes are projected to boundary plane. (C) The 2D $\alpha$-shape algorithm is applied to projected bead positions in order to determine the envelope. (D) From multiple envelopes a 3D hexahedral substrate mesh is reconstructed. Meshing with \textit{gmesh} and visualization with \textit{paraview}. Scale bar $50\ \mu$m.}
\label{femtfm_alpha_shapes_and_hexmesh}
\end{figure}

We developed a custom mesh generation program as illustrated in Fig.~\ref{femtfm_alpha_shapes_and_hexmesh}.
In Fig.~\ref{femtfm_alpha_shapes_and_hexmesh}A we show an image of the marker beads; the inset
shows the strong anisotropy of the point spread function.
First bead locations are partitioned into sections along the long axis of the wavy pattern that are separated by a partition width w.
In a second step bead positions are mapped to each section plane (normal direction pointing towards the partition direction), compare Fig.~\ref{femtfm_alpha_shapes_and_hexmesh}B. By doing this, we end up with a set of separated slices associated with 2D bead distributions. After that, we calculate the 2D hull for each bead distribution using the 2D $\alpha$-shape algorithm \cite{Edelsbrunner1983} implemented in the open-source computational geometry algorithm library (CGAL) \cite{cgal}, compare
Fig.~\ref{femtfm_alpha_shapes_and_hexmesh}C. After determination of the hull for each segmented slice we again stitch the determined contours together separated by the partition width w. We then form a 3D hexahedral mesh that approximates the substrate shape. In this particular step the program uses the open source mesh generator GMSH \cite{gmsh}. A representative result for the described procedure is depicted in Fig.~\ref{femtfm_alpha_shapes_and_hexmesh}D.

\subsection{Direct elastic problem and FEM-implementation}

\begin{figure}[t]
\centering
\includegraphics[width=0.8\textwidth]{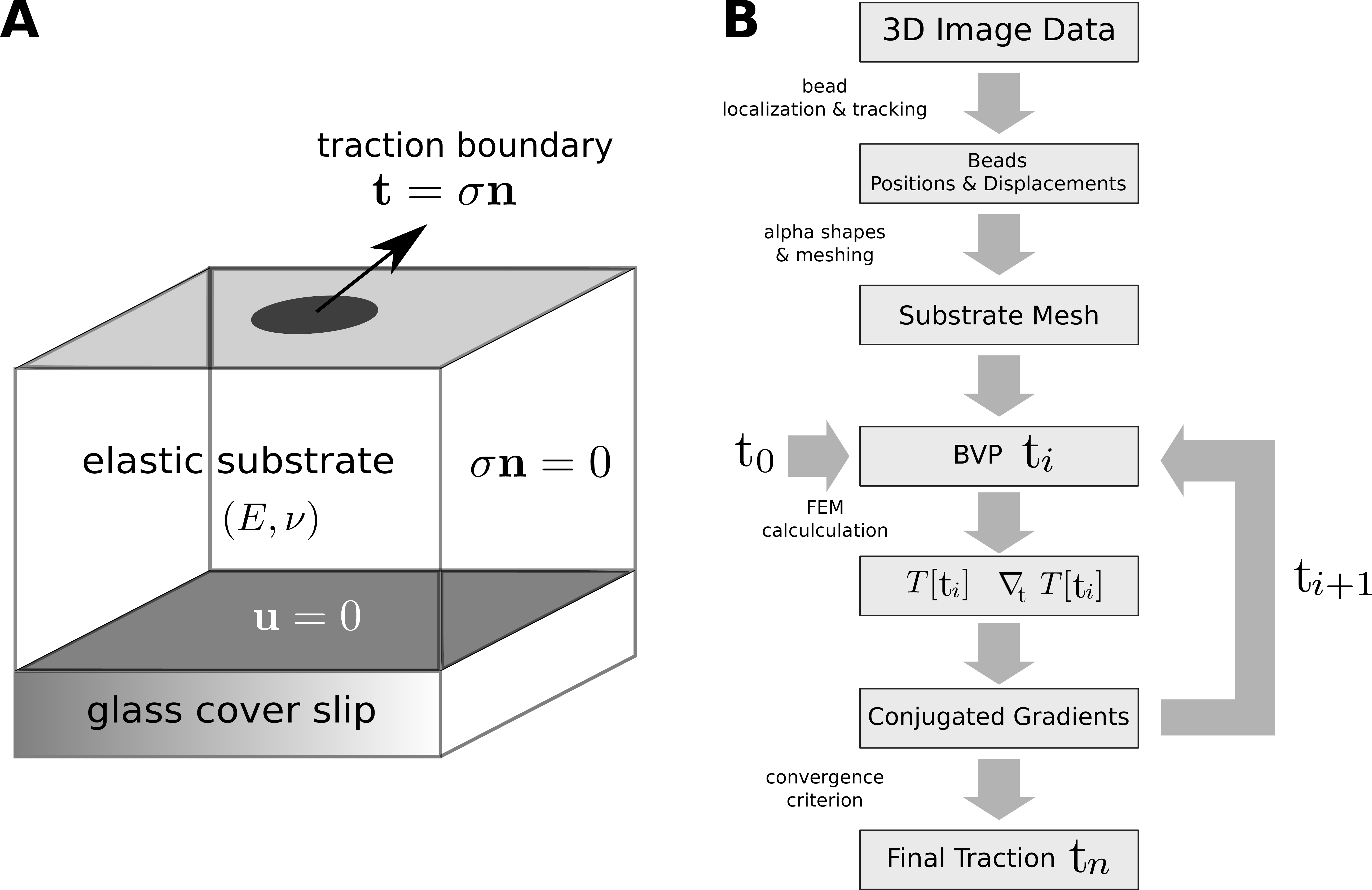}
\caption{Illustration of the direct boundary volume problem (BVP) and the work flow to solve the corresponding inverse problem. (A) Direct elastic BVP with mixed boundary conditions. The substrate elasticity is defined by Young's modulus $E$ and Poisson's ratio $\nu$. Boundary conditions: bottom - zero displacement ($\mathbf{u}=0$), (sides) - zero-stress ($\sigma \mathbf{n}=0$), top - traction field ($\mathbf{t}(\mathbf{x})=\sigma \mathbf{n}$). (B) Work flow of traction reconstruction (inverse BVP). Starting from 3D image stack raw data several steps are processed: 3D bead tracking, substrate mesh generation, FEM-calculations and numeric traction optimization.}
\label{femtfm_fig_BVP_and_reconstruction}
\end{figure}

Traction reconstruction is established by solving an inverse elastic boundary volume problem (BVP). 
We first describe the corresponding direct BVP. Because Green's functions are not known 
for the wavy shapes considered here, we use the finite element methods (FEM). In principle,
this allows us to use also non-linear formulations for large deformations and non-linear materials.
However, here we deal with linear material (PDMS) and small strains, thus the linear
formulation is sufficient. We therefore have to solve the Navier-Lam\'e equation 
for the displacement field $\mathbf{u}(\mathbf{x})$ in the computational domain $\Omega$:
\begin{equation}
	\mu \Delta \mathbf{u} + (\lambda + \mu) \nabla \nabla \cdot \mathbf{u} = 0\ .  \label{direc_BVP_first}
\end{equation}
Traction exerted by cells is applied to the top surface $\partial \Omega_{top}$ of the substrate volume in the reference state $\Omega$. Accordingly the traction field $\mathbf{t}(\mathbf{x})$ enters as stress boundary condition, $\mathbf{t}(\mathbf{x}) = \sigma(\mathbf{x}) \mathbf{n(\mathbf{x})}$, where $\mathbf{n}$ is the normal vector of the unit surface element and $\sigma(\mathbf{x})$ the substrate Cauchy stress tensor. As illustrated in Fig.~\ref{femtfm_fig_BVP_and_reconstruction}A, we choose appropriate mixed BCs at the remaining parts of the boundary arriving at a well-defined BVP. At the bottom surface we require zero displacement, $\mathbf{u}(\mathbf{x})=0$, due to rigid coupling between the soft elastomeric substrate and the underlying rigid glass coverslip. Since the used  mesh represents a cutout of the substrate, which is largely extended in lateral directions, proper boundary conditions at the side surfaces are applied. In case of an infinite half-space, an appropriate boundary condition for the side surfaces of the cutout would be a counter stress of the same magnitude. This however would lead to an undesired recursive problem. Here we use vanishing stress boundary conditions, $\sigma \mathbf{n} = 0$, for sufficient large cutouts. Based on the knowledge that the displacement field monotonically decays at least as $1/r$, the influence of boundary conditions can be neglected for sufficient large cutouts. Therefore, we occasionally extend the substrate mesh in lateral directions to ensure the repression of boundary effects. 

For a given traction pattern $\mathbf{t}(\mathbf{x})$, the direct BVP formulated in  Eq.~\ref{direc_BVP_first}
is now solved by means of the FEM as typically applied to elastic problems \cite{Braess_book,Ameen_book,Bonet1997}.
For this purpose, we transform the elastic equations into the weak form and use linear shape functions
on a hexahedral lattice to arrive at an algebraic problem (the mathematical details are provided as supplementary material). 
The algebraic equations are solved with the conjugated gradient (CG) method. Alternatively, it is also possible to directly invert them
by means of e.g. Gauss-elimination. The achieved solution then can be used to interpolate the displacement to any
position within the domain $\Omega$. For the implementation of our FEM-approach, we used the FEM C++ library {\it Deal.II}
\cite{Bangerth2007}. It provides the essential set of features to achieve a FEM-calculation, for instances, managing local and global DoF indices, matrix manipulation feature, Gauss-quadrature, coordinate transformations and solvers for the linear algebraic system.

\subsection{Inverse elastic problem}

The inversion of the direct BVP is in general ill-posed, which implies that no unique or sufficiently stable solution exists.
Therefore regularization methods have to be used to render the inverse problem stable.
According to Tikhonov and Arsenin \cite{Tikhonov1977}, the regularized inverse problems can be formulated as a minimization problem and in TFM the corresponding Tikhonov functional $T[\mathbf{t}(\mathbf{x}),\lambda]$ has the general form:
\begin{equation}
      T[\mathbf{t} (\mathbf{x}'),\lambda] = L[\mathbf{u}(\mathbf{x}, \mathbf{t} (\mathbf{x}'))] + R[\mathbf{t}(\mathbf{x}'),\lambda] \label{eq:general_target_functional}
\end{equation}
Here, $\mathbf{x} \in \Omega$ and $\mathbf{x}' \in \partial \Omega_{top}$. $L[\mathbf{u} (\mathbf{x}, \mathbf{t} (\mathbf{x}'))]$ represents an error estimate that assigns a value to differences between calculated and measured displacements (the \textit{residue}). The larger the deviation, the larger is the scalar value the estimate returns. The second term $R[\mathbf{t} (\mathbf{x}')]$ is a penalty functional introduced to recover a well-posed solution (the \textit{regularization term}) \cite{Vogel2002}.
  
We discretized the traction field $\mathbf{t}(\mathbf{x})$ according to the FEM mesh in order to set up a finite space of parameters. For this purpose, we use interpolation based on shape functions. Hence, the entire field is characterized by a set of fixed point values $t_i$, with $i \in \{1 ... N_t \}$, which represent the degrees of freedom (DoFs) for the considered optimization problem. Since we use linear shape functions, fixed point values are associated with nodal positions on the top surface of the FEM mesh $\partial \Omega_{top}$. Thus, the total number of optimization parameters $N_{t}$ is determined by the number of surface mesh vertices $N_v^{top}$, and the number of space dimensions ($N_d=3$), $N_{t}= 3*N_v^{top}$. The discretized version of the Tikhonov functional then reads
$T[\mathbf{t} (\mathbf{x}'),\lambda] = L(\mathbf{u} (\mathbf{x}, \{t_1, ..., t_{N_{t}} \})) + R(\{t_1, ..., t_{N_{t}} \},\lambda)$.

We next need to define the form of the residue and the regularization term in Eq.~\ref{eq:general_target_functional}.
Standard TFM uses the least square estimate $L(\mathbf{u} (\mathbf{x}, \mathbf{t} (\mathbf{x}')))= \sum \limits_{i=1}^{N_{beads}} \|\mathbf{u} (\mathbf{x}, \mathbf{t} (\mathbf{x}')) - \mathbf{u}^{exp}_i \|^2$ to measure deviations between measured and computed displacements. Additionally, most methods use $0$th order Tikhonov regularization $R[\mathbf{t} (\mathbf{x}'),\lambda] = \lambda \int_{\partial \Omega_{top}} \|\mathbf{t}(\mathbf{x}') \|^2~dA$. This enforces a smooth traction solution by penalizing the amount of total force and thus represents the most simple approach to repress noise-induced fluctuations \cite{Dembo1999,Schwarz2002,Sabass2008}. Thus
in the standard approach, both terms employ the L2-norm. Here we write a more general form:
\begin{equation}
      T[\{t_1, ..., t_{N_{t}} \},\lambda] = \sum \limits_{i=1}^{N_{beads}} \| \mathbf{u}_{FEM} (\mathbf{x}_i; \{t_1, ..., t_{N_{t}} \}) - \mathbf{u}_{exp}(\mathbf{x}_i)  \|^{p_L} + \lambda \sum \limits_{j=1}^{N_{t}}  \| t_j \|^{p_R}\ . \label{eq:standard_discrete_functional}
\end{equation}
The choices $p = 1$ and $p = 2$ correspond to the L1- and L2-norms, respectively. 
Below we will always use the standard choice $p_R = 2$ for the regularization term.
For the residue, we will first use the standard choice $p_L = 2$, which corresponds to the least square estimator.
From a statistical point of view, the least square estimator can be derived as the maximum likelihood estimate (MLE) for Gaussian distributed errors \cite{Maronna2006}. Later we will argue that for our case, $p_L = 1$ (L1-norm for the residue) is a better
choice in our case, because it deals better with the outliers in our experimental data.

The form written in Eq.~\ref{eq:standard_discrete_functional} implicitly assumes an isotropic error (same error distribution in all spatial directions). In our case, this assumption is not satisfied anymore, due to a reduced resolution in $z$-direction, compare Fig.~\ref{femtfm_alpha_shapes_and_hexmesh}A. Therefore, we introduce a scaling procedure that weights the estimate contribution due to their relative accuracy. In detail, the $z$-contribution is weightened with a factor $1/3$ compare to the $x$- and $y$-contributions, which are assumed to have the same weight.

Due to repeated time consuming FEM-calculations of the direct BVP during minimization of the Tikhonov-functional, efficient computation is a key issue in solving the inverse problem. The overall computation time depends in main parts on the number of traction DoFs $N_t$. Hence, a major objective was to achieve the best possible local accuracy for a given number of DoFs $N_t$. In order to achieve this demand we used predefined mesh refinement and adaptive local mesh refinement (h-refinement). Mesh refinement is employed by dividing selectively volume elements into smaller elements, which effectively increases the density of DoFs. The idea behind it is to use local variations of the mesh size to concentrate DoFs at regions with higher levels of $\mathbf{t}$. Other region far away from the traction sources remain coarser at the same time. For the sake of completeness, we want to mention the alternative of local polynomial refinement, which describes local variation of the polynomial degree of used shape functions, called p-refinement. Also combinations of both refinement types in terms of hp-refinement schemes are conceivable. However, as h-refinement and p-refinement have similar effects to the local resolution, we here consider h-refinement only. In practice, the h-refinement in our program is done by an adaptive scheme. Based on reconstruction results on a coarser mesh, it is decided by global thresholding whether an element gets refined. Afterwards the reconstruction process is restarted with the interpolated traction field obtained before. This process is repeated until a desired local resolution is achieved. Alternatively, we implemented also predefined local refinement based on thresholding of the measured displacement field. For the calculations in this work, we keep at local adaptive local mesh refinement schemes. This procedure is tested below by reconstruction of simulated data.

\subsection{Optimization Procedure}

The core module of the FEM traction reconstruction program is the implemented multidimensional parameter optimization procedure based on the conjugated gradient (CG) method. The implementation follows essentially the Fletcher-Reeves variant of this algorithm as described in \cite{NumericalRecipes2007}. Different from the standard implementation, we parallelized most parts of the procedure, like the numeric gradient calculation $\nabla_t T[t_i]$ and the subsequent line minimization. Parallel computing was realized by using the Message Passing Interface (MPI), which is suitable for large scale distributed computing on computer clusters or sheared memory systems. Although the CG method is a common tool in the field of inverse problems, e.g. \cite{Lukyanenko2012}, we tested other optimization methods as well,
namely gradient-less downhill-simplex optimization, simple steepest descent and heuristic Monte-Carlo optimization with simulated annealing (all described in \cite{NumericalRecipes2007}). We found that the CG method led to the shortest computation times and excellent convergence. Fig.~\ref{femtfm_fig_BVP_and_reconstruction}B shows a schematic representation of our complete workflow.

\section{Results}

\subsection{Method validation}
\label{sec:femtfm_model_simulations}

\begin{figure}[h]
\centering
\includegraphics[width=1\textwidth]{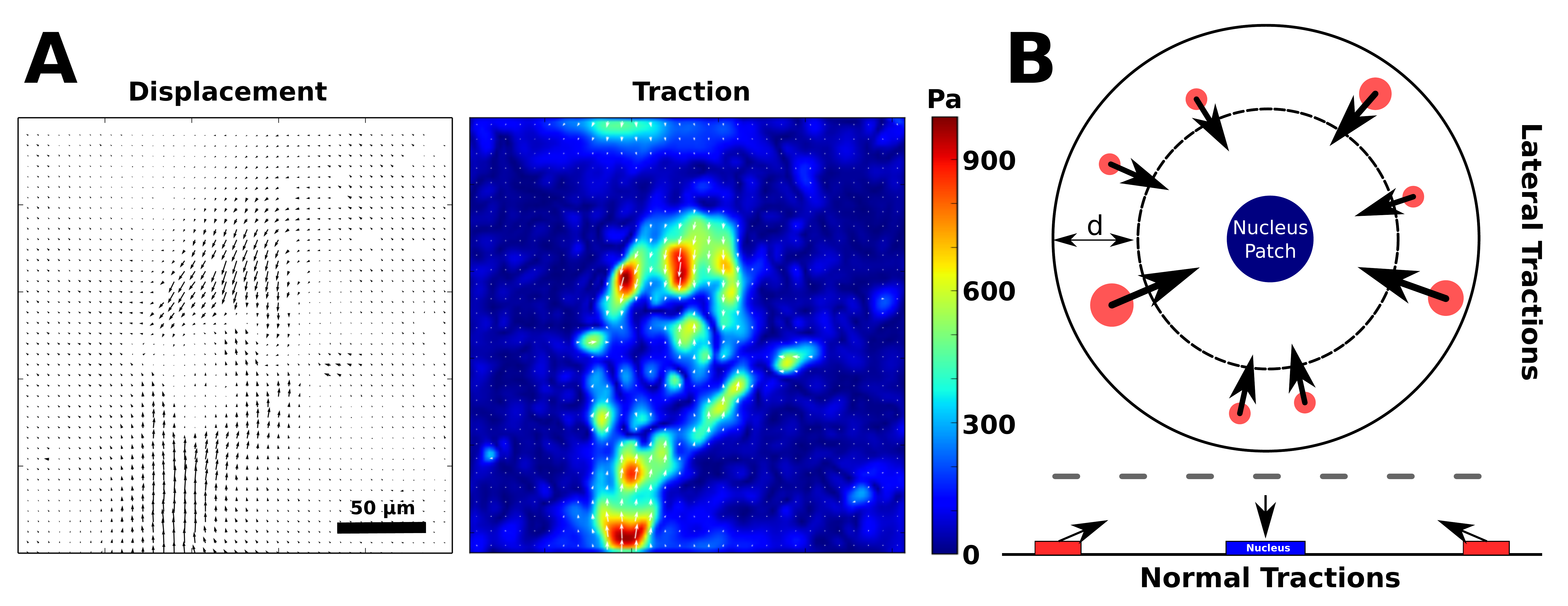}
\caption{Simulation of traction patterns. (A) Experimental example: cardiac myofibroblast on a $15$ kPa PDMS substrate. Substrate displacement field and
2D traction forces obtained from PIV and regularized Fourier transform traction cytometry (reg-FTTC) \cite{Sabass2008}. The cell shows
a typical dipole pattern with strong inward directed forces at the cell periphery.
(B) Illustration of traction pattern simulation. Traction patches are distributed within a ring-shaped area close to the cell edge
with normal components that are balanced by a central patch that is consequently pushed into the substrate.
Traction magnitude in lateral direction is set to a constant value of $3$ kPa. }
\label{femtfm_fig_cell_like_simulation}
\end{figure}

As reported earlier, for 2D TFM our implementation of FEM-TFM gives equivalent results as the standard
approach with FTTC \cite{Schwarz2015}. In this case, FTTC is much more efficient and faster.
Here however we want to address 3D TFM for wavy substrates in which a Green's function is not known, making
FEM-TFM the most reasonable approach. In order to validate our method for a 3D situation,
we first reconstructed simulated data for planar substrates
with both tangential and normal components. We generated these distributions by a simple set of rules that mimic typical cellular force distributions. 
Fig.~\ref{femtfm_fig_cell_like_simulation}A shows a typical FTTC traction force pattern
for a cardiac myofibroblast cultured on a 15 kPa PDMS substrate. This cell shows the typical dipole
pattern that has been suggested as a minimal model for a contractile cell \cite{schwarz_physics_2013}.
Fig.~\ref{femtfm_fig_cell_like_simulation}B illustrates the rules we use to generate artificial traction patterns.
A cell is modeled as a circle with traction patches located only in a peripheral annulus of
width $d$. Each of these patches carries a tangential traction stress of $f=3$ kPa which
is a typical value found for cells \cite{Balaban2001,Fu2010}.
However, previous studies also reported appreciable cellular traction stress in normal directions \cite{Hur2009,Delanoe-Ayari2010,kristal2013metastatic}.
They found a typical pull-push pattern, such that the cell is pulling up at the periphery and pushing down
with the cell body. The upward forces might arise from actin fibers being anchored
at the dorsal side of the cell or at the upper side of the nucleus, while the downward force might be 
the reaction force localized by the large and relatively stiff nucleus \cite{DvirWeihs2015}. Here we include
these normal forces in our simulations by adding force in positive z-direction to our
adhesion patches, that are counterbalanced by an extra traction patch located at the cell center.
Together, these rules allow us to generate realistic traction distributions that satisfy basic properties of a
typical cell induced 3D traction pattern.

The displacement fields resulting from these simulated traction patterns were calculated using 
the direct BVP introduced above (Eq.~\ref{direc_BVP_first}) with our FEM-implementation.
From the resulting displacement field, we sampled $N_{bead}$ random displacements. To account for uncertainties due to the contribution of experimental noise, we introduced additive random displacement errors $\mathbf{u}_{err}$ that modify each displacement component by a random value. Subsequently a simulated bead displacement is expressed by $\mathbf{u}_{bead} = \mathbf{u}_{FEM} + \mathbf{u}_{err}$. The simulated data allows us to validate and characterize features and limits of our method under well-controlled conditions.

\begin{figure}[t]
\centering
\includegraphics[width=1\textwidth]{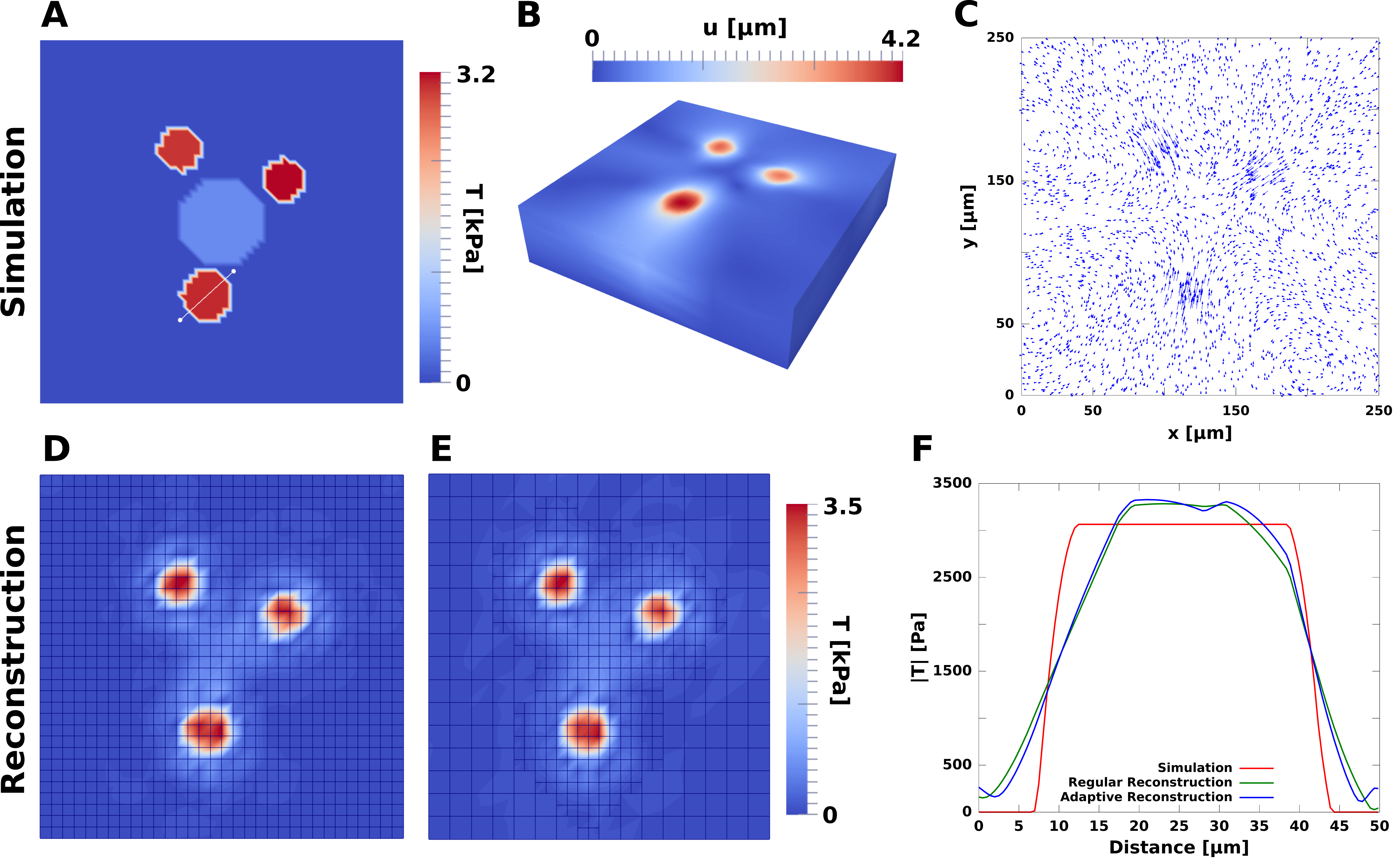}
\caption{Reconstruction of simulated traction forces with the finite element method (FEM-TFM). (A)-(C) Simulation of bead displacement. 
Simulated bead density $n_{bead}=0.048$ beads / $\mu m^2$, which corresponds to a bead number of $N_{bead}=3.000$.
(D) Traction reconstructed with FEM on regular mesh (3.267 degrees of freedom, DoFs).
(E) Reconstruction on locally h-refined mesh (1.848 DoFs) works equally well.
(F) Comparison of simulation and reconstruction along a traction profile through a force patch (white line in (A)).}
\label{femtfm_fig_adaptive_traction_reconstruction}
\end{figure}

In Fig.~\ref{femtfm_fig_adaptive_traction_reconstruction}A-C we show the results of a typical simulation without
noise ($\mathbf{u}_{err}=0$). Figs. \ref{femtfm_fig_adaptive_traction_reconstruction}D and E depict a comparison of reconstruction results on a homogeneous and a adaptively refined mesh, respectively. For the adaptive reconstruction, we started with a two times larger mesh size compared to the homogeneous mesh. After one refinement step, we subsequently achieve the same mesh size at refined regions, which makes results comparable. Both results show an excellent agreement and reconstruct correctly the original traction pattern. The depicted mesh topology shows that the adaptive refinement automatically adjusts the local mesh size to regions of accumulated traction stress, where we accordingly establish higher local accuracy. By doing this we saved $47\%$ of Degrees of Freedom (DoFs) which has a positive effect on the computation time. Fig.~\ref{femtfm_fig_adaptive_traction_reconstruction}F shows the traction profile through a reconstructed traction patch. Both reconstructions lead to a slightly smoothed profile compared to the original. This reduced resolution of a sharp edge can be explained by a limiting bead density (in our simulation $n_{bead}=0.048 / \mu m^2$) according to the sampling theorem by Nyquist and Shannon. From the simulation results we conclude that adaptive local mesh refinement has no detrimental effect on the traction reconstruction accuracy. However, it clearly needs much less DoFs and therefore is much more computer time efficient.

When solving an inverse problem the stability of the solution strongly depends on the uncertainties in the provided data \cite{Vogel2002}. In particular,
the problem might be ill-posed due to the effect of noise. This has been explicitly shown for the case of traction reconstruction \cite{Schwarz2002}. As mentioned in the introduction, experimental conditions limit the resolution measured displacement fields. Main reasons here are limited optical resolution of the microscope and errors in the bead tracking procedures. The first issue can be treated as Gaussian shaped errors that limits the spatial localization of beads \cite{Plotnikov2014}. The latter is less relevant in the planar 2D case since state of the art 2D tracking methods are very accurate with eventually negligible error rates. However, bead localization and tracking in 3D is by far more challenging, even with modern microscopic setups. This is due to anisotropic optical resolution that produce a notably inferior accuracy in z-direction compared to the corresponding lateral directions. There already exist improved setups like dual objective super resolution microscopy techniques \cite{Shtengel2009,Xu2012,Aquino2011}, however, these are not standard and rarely available and because of this no application in TFM has been shown so far. Therefore, we aimed at improving bead tracking in 3D and on adapting the traction reconstruction method to data with anisotropic optical resolution. In fact we realized that due to the substrate topography we have to deal with a locally varying point spread function. That directly effected the tracking of bead movements when using image cross correlation techniques. Hence, we applied a more robust single bead tracking method than done in standard TFM,
using feature vectors introduced earlier for 3D TFM for the same reasons \cite{Legant2010}. By using this technique we obtained a significantly improved displacement field, however, the data showed strong and inhomogeneous drift which could not be corrected by a single drift vector (evaluated in one focal plane). This effect can be explained by relatively long image stack acquisition times of approximately $30$ min when recording $\approx 100$ images per stack. Due to this long acquisition time we observed non-monotonic drift when comparing two image stacks. In order to solve this problem, we applied drift correction for each individual slice, which successfully canceled out most of the drift. Nevertheless, the derived displacement data showed anisotropic deviations and occasionally higher densities of outliers.

\begin{figure}[t]
\centering
\includegraphics[width=1\textwidth]{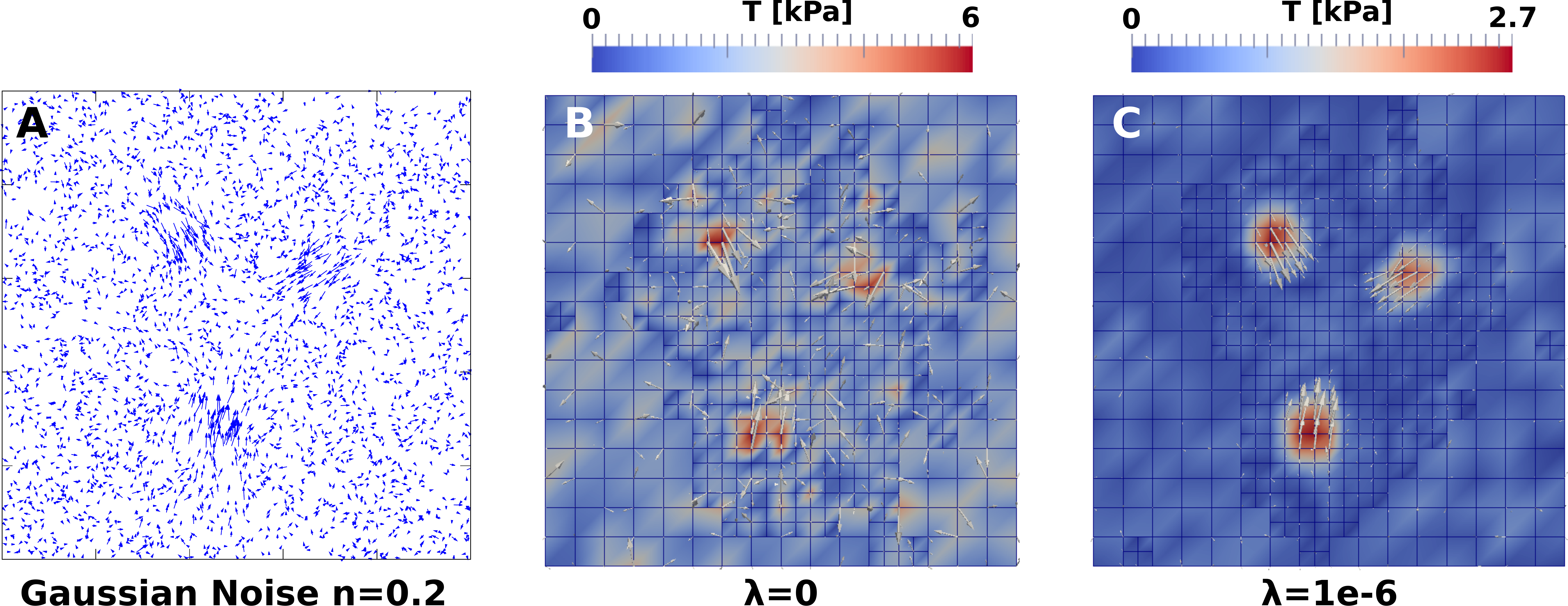}
\caption{Effect of Gaussian noise and the need for regularization. (A) Simulated displacement field 
from Fig.~\ref{femtfm_fig_adaptive_traction_reconstruction} with Gaussian noise added
(noise level $20\%$). (B) Reconstruction without regularization does not
give a clear traction pattern. (C) Reconstruction with a regularization parameter $\lambda=1e{-6}$, selected by
generalized cross validation (GCV),
reconstructs the original traction pattern, however, with a reduced maximal traction magnitude of
$\sim 11\%$ compared to the simulated pattern.}
\label{femtfm_fig_gaussian_noise_reconstruction}
\end{figure}

Fig.~\ref{femtfm_fig_gaussian_noise_reconstruction}A shows the displacement field used in Fig.~\ref{femtfm_fig_adaptive_traction_reconstruction}
with Gaussian noise of strength $20\%$ (measured with respect to maximal displacement).
The force reconstruction without regularization is shown as
Fig.~\ref{femtfm_fig_gaussian_noise_reconstruction}B and clearly is not very good. 
Therefore we next used $0$th order Tikhonov regularization ($R=\lambda \|\mathbf{t} \|^2$) to find an approximated reconstruction solution. Here, we determine an optimal regularization parameter by using generalized cross validation (GCV) \cite{Hansen1992}. The optimal value of $\lambda$ depends on the noise level and for the given example, we determined $\lambda=1e{-6}$. The regularization improves the result significantly, as shown in Fig.~\ref{femtfm_fig_gaussian_noise_reconstruction}C. The penalization of total force induced by $0$th Tikhonov regularization, however, led to traction underestimation and edge smoothing.

Next, we simulated the presence of outliers. We drew the additive random displacement error $\mathbf{u}_{err}$ for each bead either from a Gaussian distribution $N(\sigma_{noise})$ or from a box-shaped distribution $H_i(w_i)$ limited by the box width $w$. We decided from which distribution an error contribution is drawn by calculating an equally distributed random variable $\mathbb{X}$ over the interval $[0,1]$. If $\mathbb{X}<=\epsilon$, we draw $u_{err}$ from the boxed-shaped distribution, corresponding to an outlier. In case of $\mathbb{X}>\epsilon$, we calculated a Gaussian distribution $u_{err}$. Here, $\epsilon$ corresponds to the fraction of outliers. In general, there are two possible approaches to diminish the effect of outliers for the reconstruction. One is to filter out outliers based on predefined criteria. This demands to set up appropriate filter limits. Moreover, such a procedure might overlook valuable information in the data. As a second approach one can use robust estimates for the optimization process. We will show in the following that reconstruction can be improved, when the least square estimate in the Tikhonov functional gets replaced by a robust measure more insensitive against data outliers. We tested different robust maximum likelihood estimates known from optimization theory \cite{Maronna2006}.

The simplest robust estimate uses the L1-norm, $\rho(x)= |x/\sigma|$. As the L2-estimate described above, it can be derived analytically from the maximum likelihood assumption, in this case based on a Laplace distribution $f(x)=\frac{1}{2\sigma}\exp \left(-|x|/\sigma \right)$. The essential advantage of this estimate is its weighting of deviations, which compared to the L2 is less sensitive to outlier contributions. As alternative estimates, we implemented also Huber functions or biweighting functions \cite{Maronna2006}. They require an additional cutoff parameter $k$ to characterize transition feature of the effective deviation weighting. This indeed requires a good guess about expected outlier strength.

\begin{figure}[t]
\centering
\includegraphics[width=1\textwidth]{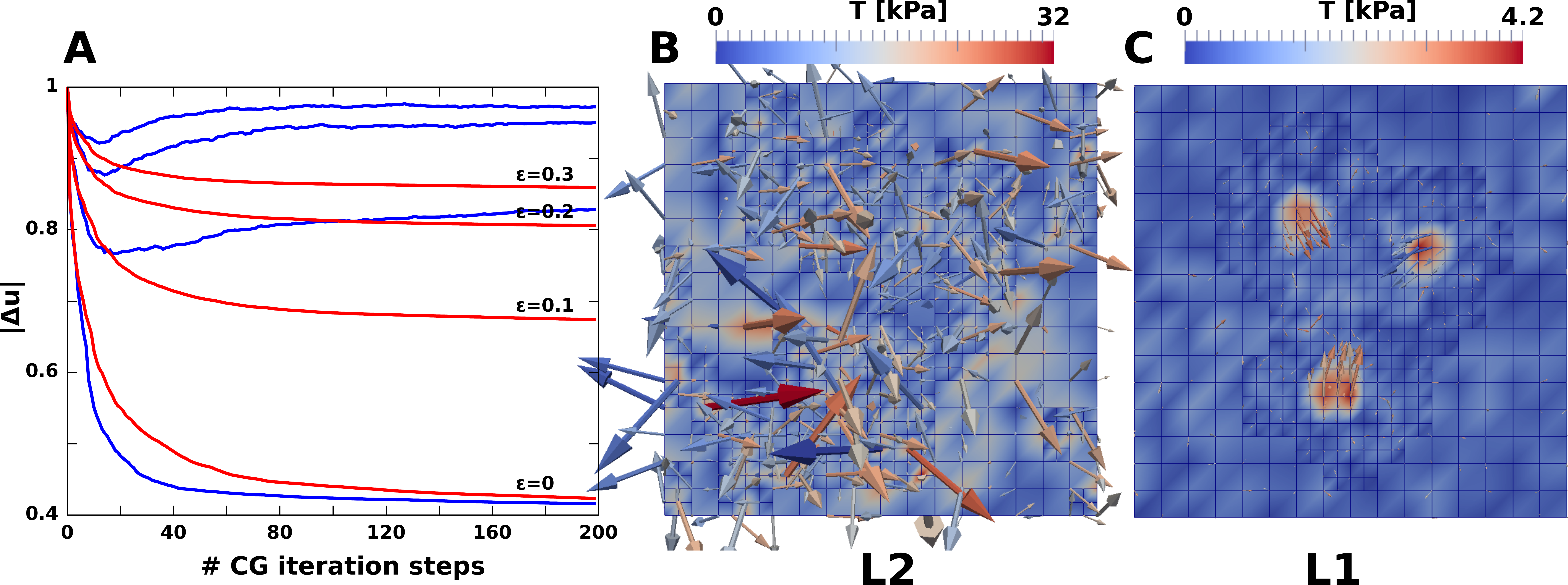}
\caption{Reconstruction of data with mixed error, including outliers. (A) Mean relative deviation plotted over the number of optimization steps for different outlier fractions $\epsilon=0,0.1,0.2,0.3$ and for two different estimates, L2 (blue) and L1 (red). (B) L2-based reconstruction for outlier fraction $\epsilon=0.3$. (C) L1-based reconstruction for outlier fraction $\epsilon=0.3$. For all simulations the Gaussian noise level was
$10\%$.}
\label{femtfm_fig_reconstruction_outliers}
\end{figure}

Fig.~\ref{femtfm_fig_reconstruction_outliers} depicts the influence of the used estimate on the quality of the reconstructed traction field. The simulation study considered an isotropic Gaussian based noise of $n_{noise}=0.1$ and a isotropic box shaped distribution with $w=10\ \mu m$. We investigated the convergence behavior for the optimization achieved by L1- and L2-estimates with respect the outlier fractions $\epsilon=0,0.1,0.2,0.3$. To compare the results, we introduced the relative displacement deviation $|\Delta u|=\sum \limits_{i=0}^{n} |u_{i}^{exp}-u_{i}^{FEM}|/|u_i^{exp}|$. For $\epsilon=0$ (no outliers), both estimates converge to a similar result, however, the L1 converges slower. If we chose a non-zero fraction of outliers $\epsilon>0$, the convergence dynamics of the optimization procedure starts to differ between L2 and L1. The L1 shows still a monotonic decreasing $|\Delta u|$, while in case of the L2 the curve starts to increase again saturating into a different solution. Corresponding traction field solutions for $\epsilon=0.3$ are depicted in Fig.~\ref{femtfm_fig_reconstruction_outliers}. In case of using the L2, the reconstruction is strongly influenced by the fraction of outliers. In contrast, the L1-reconstruction leads to satisfying results comparable to the target field.

\begin{figure}[t]
\centering
\includegraphics[width=0.6\textwidth]{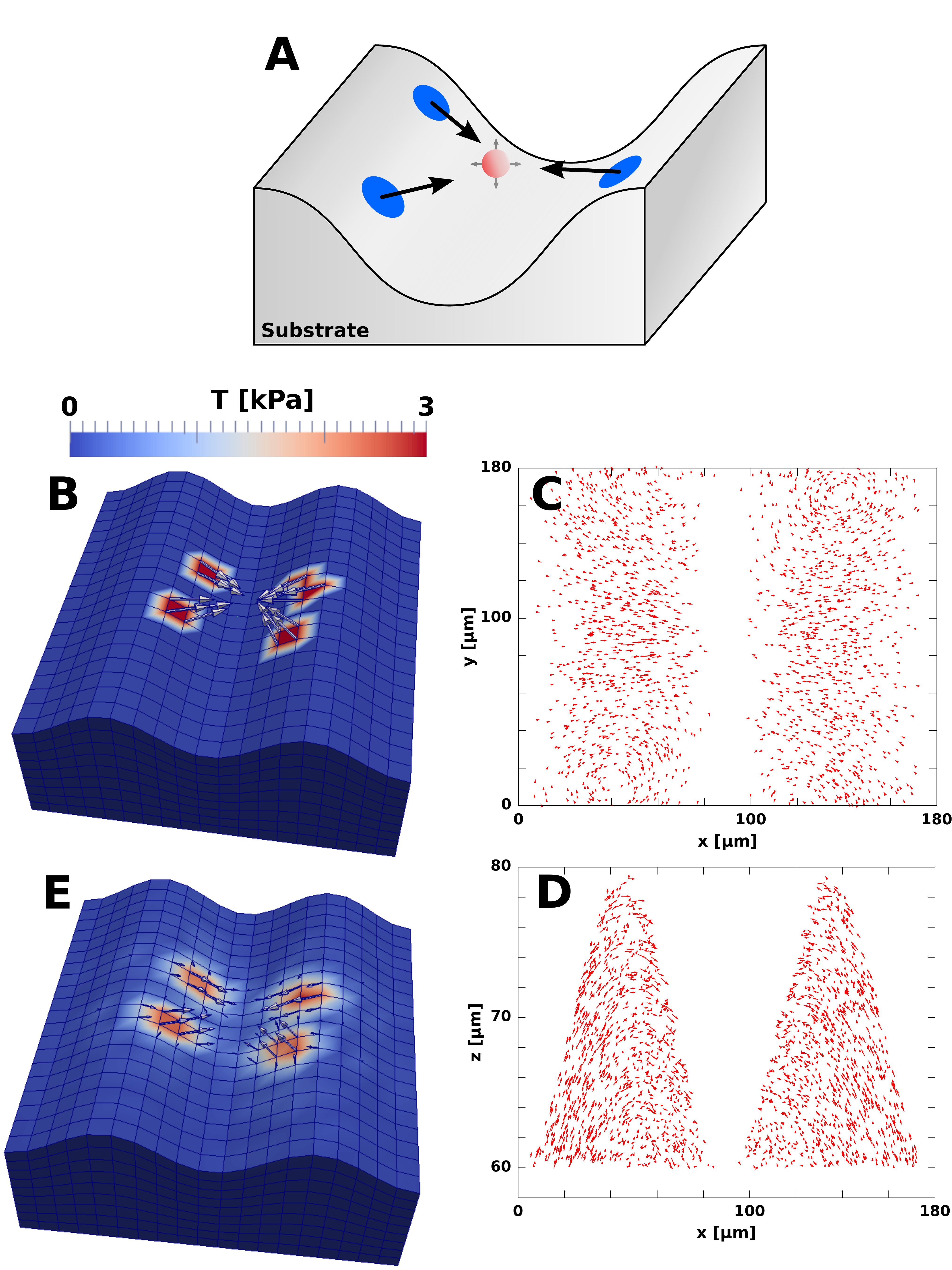}
\caption{Traction reconstruction simulated on a sinusoidal substrate topography. (A) Illustration of the simulation scheme. All forces point to a virtual point (red) between the two ridges to ensure force balance. (B) Simulated traction field with four traction patches $f=3$ kPa. Simulated displacement field with bead volume density $n^{V}_{bead}=0.003$ beads / $\mu m^3$ ($N_{bead}=2000$), (C) projected to the xy-plane, and (D) projected to the xz-plane. (E) The achieved reconstruction is very close to the original traction pattern.}
\label{femtfm_fig_cellb_reconstruction}
\end{figure}

After having established a successful approach for planar substrates with both tangential and normal forces,
we finally simulated 3D FEM-TFM with wavy substrates as shown schematically in Fig.~\ref{femtfm_fig_cellb_reconstruction}A.
We first distributed random traction patches at both flanks of the sinusoidal shape, while the traction vectors point at a common virtual point. In a second step, all traction directions got reoriented due to movement of the virtual point in 3D, until a balance of overall force was achieved.
Fig.~\ref{femtfm_fig_cellb_reconstruction}B and C depict an exemplary simulation result. It shows a traction field with four distinct traction patches with a force density of $f=3$ kPa. The corresponding simulated bead displacements are depicted in C and D as of projections to the xy-plane and xz-planes, respectively. Compared to the planar case, the displacements in z-direction became more prominent, see Fig.~\ref{femtfm_fig_cellb_reconstruction}D. The resulting reconstruction shown in Fig.~\ref{femtfm_fig_cellb_reconstruction}E
shows similar good agreement as for the planar case, thereby validating our method also for the experimental
setup to be studied.

\subsection{Morphology and traction forces of cardiac myofibroblasts on wavy substrates}
\label{sec:femtfm_results}

\begin{figure}[t]
\centering
\includegraphics[width=1\textwidth]{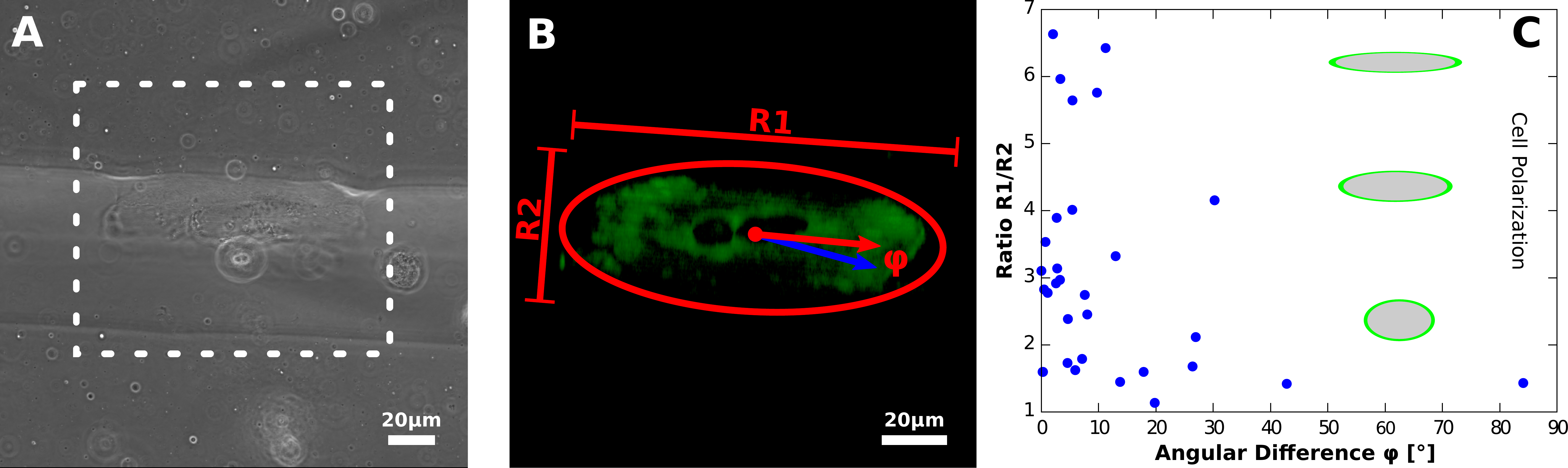}
\caption{Cell polarization and orientation. (A) Phase contrast image of a cardiac myofibroblast aligned with substrate topography. (B) Illustration of ellipse fitting to analyze cell orientation and shape in terms of ratio between lengths of semi-major R1 and semi-minor R2 from images of vinculin-transfected cells. The angular difference is defined as angle $\varphi$ between cell orientation (red arrow) and substrate orientation (blue arrow). (C) Cell shape plotted versus the angular difference $\varphi$.}
\label{femtfm_fig_cell_morphology}
\end{figure}

We finally investigated myofibroblasts adhered to elastic wavy PDMS-substrates with a focus on their morphology and traction forces. 
For our context, the choice of PDMS has several advantages over the PAA-system often employed for planar 2D TFM.
In particular, it does not swell due to water uptake, it has superior optical properties, and it can be 
reproducibly molded into wave structures on a microscopic scale of several tens of micrometers.
We first quantified the effect of substrate topography on cell morphology and cytoskeletal polarization. For this purpose we evaluated a data set of stack images of $30$ cells adhered to waves of $\sim 70\ \mu m$ width and $\sim 30\ \mu m$ height. A representative
phase contrast image is shown in Fig.~\ref{femtfm_fig_cell_morphology}A. Such images
were used to determine the orientation of the topographic features. The cells were
fluorescently stained for cell adhesion-related molecules to measure their internal
organization. We determined the direction and degree of cell polarization by fitting a 3D ellipse to the image and subsequently evaluated the orientation and eccentricity of the ellipse, compare Fig.~\ref{femtfm_fig_cell_morphology}B. 
We finally correlated the two measures. Fig.~\ref{femtfm_fig_cell_morphology}C shows the
eccentricity as a function of the 
angle between substrate and cell orientations. The plot confirms that cells tend to align with the long axis of the substrate. 2/3 of the cells show a ratio $R1/R2$ (semimajor over semiminor) larger than 2 and $\sim 80\%$ are aligned with the wave within the angular range of $0$ to $15^\circ$. This indicates that even the cells that are close to round have sensed the topographic features of the wavy substrate.
From our ellipse analysis we further found that they tend to adhere most often to the wave flanks and less to the valleys or hill tops of the height-modulated structures.

\begin{figure}[t]
\centering
\includegraphics[width=1\textwidth]{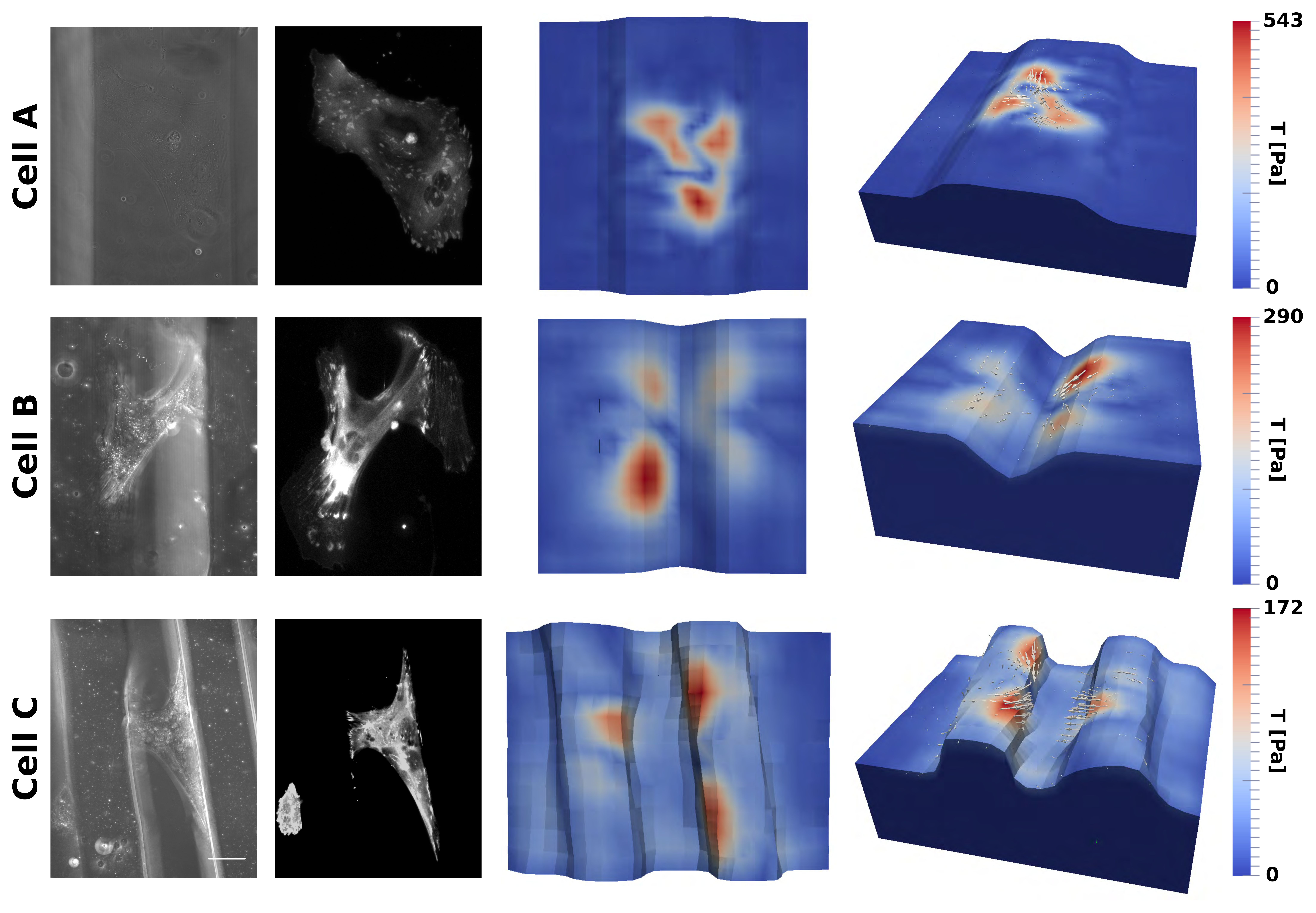}
\caption{Traction reconstruction for cardiac fibroblasts cultivated on three different topographies. From left to right we show phase contrast image, VASP-GFP fluorescence image, top view and 3D view of reconstructed traction field, respectively. Scale bar $50 \mu$m.}
\label{femtfm_fig_result_reconstruction}
\end{figure}

In addition to the morphological study we applied the FEM traction reconstruction to an exemplary set of cells. Here, we wanted to investigate the influence of environmental topography on cellular contractility. The long acquisition time for displacement data however
made it very difficult to conduct a full statistical analysis and therefore in Fig.~\ref{femtfm_fig_result_reconstruction} we only
show a few typical examples of cardiac myofibroblasts cultured on wavy substrates with a Young's modulus  of $E=15$ kPa.
In Fig.~\ref{femtfm_fig_result_reconstruction} two different topographies are used, micro-grooves and wave shapes.
For the substrates in cases A/B/C we used 70/142/91 slices with z-distance between two slices of 0.55/0.39/0.30 $\mu m$.
We tracked 6205/9614/13613 beads, with an average distance between beads
of 5.40/5.91/2.90 $\mu m$. For the reconstruction, we used the adaptive mesh refinement approach and started to calculate a traction solution on a coarser mesh ($\sim 50$ optimization steps). Afterwards the algorithm refined $\sim 40\%$ of the computational grid cells at the surface based on obtained traction magnitudes. We subsequently achieved local mesh sizes of $d_{m}=7/10/6 ~\mu m$ for the corresponding cells A/B/C (see Fig.~\ref{femtfm_fig_result_reconstruction}). We further utilized an L1-estimate for the residue and a 0th order Tikhonov regularization with L2-norm (same regularization parameter for all data). As explained above, the
$z$-direction is weightened only with a factor of $1/3$ due to its poor resolution compared to the two lateral directions.

Cell A was adhered on top a slightly curved surface between two grooves (dimensions: $h=25~\mu m$, $w=120~\mu m$). Compared to the others it shows the largest spread area and a roughly homogeneous distribution of adhesion sites (observed in vasodilator-stimulated phosphoprotein (VASP) fluorescence image). For this cell, we obtained the largest maximal traction as $T_{max}=543$ Pa. Cell B occupied the edge region of two opposite groove flanks (groove dimensions: $h=30\ \mu m$, $w=~70\ \mu m$). FAs were concentrated to limited region at the edge and the cell spanned the groove with actin SFs bridging free space. The maximal traction was $T_{max}=290$ Pa. Strikingly the traction vectors revealed that most of the forces were balanced along the edges and not to the opposite groove flanks. Cell C spanned the gap between two wave shaped hills (wave shape: $h=~30\ \mu m$, $w=25\ \mu m$ with separation distance $d=60\ \mu m$). FAs were more concentrated at vertical parts of the edge region with a similar total area compared to Cell B. We derived the lowest maximal traction $T_{max}=172$ Pa for this cell. As for cell B, however, most of the forces seemed to be balanced along the right wave instead of across the gap. Therefore we conclude that
even if cells span two neighboring hills, they still tend to balance their forces along the ridges, in agreement with the
earlier observation that their cytoskeleton is in average organized in this direction.

\section{Discussion}

Here we have described experimental and theoretical methods to reconstruct cellular traction patterns of cells adhering to non-planar (wavy) substrates. This involves novel micromolding techniques to prepare curved elastic substrates with embedded fluorescent beads, improved
image processing procedures for bead tracking and a completely novel 3D TFM workflow to achieve traction reconstruction utilizing FEM and optimization procedures. We successfully checked the validability of our method by first reconstructing simulated data considering different experimental conditions. 

As shown in the method validation part, the presence of outliers plays a crucial role in proper traction force reconstruction for our 3D data on topographic substrates. By replacing the typically used L2- by an L1-error estimate, we demonstrated a way to prevent problems in traction reconstruction originated from outliers. The L1-estimate is a much more robust measure in the presence of outliers in noisy data. This can be essentially traced back to a reduced weighting of outliers (linear weighting) during the reconstruction process. Compared to other known robust estimates, the L1-norm needs no additionally cutoff parameters and converges to the L2-solution for outlier-free data with Gaussian error distributions \cite{Maronna2006}. We note that this approach aims at improving traction field solutions for non-smooth and noisy 3D data. It has no essential influence on the regularization scheme. Here we stay with the standard L2-norm, which seems appropriate 
for our case of dense marker beads. Recently, it has been suggested that the L1-norm is favorable for
the regularization term in the case of high resolution 2D TFM \cite{han2015traction}. Although we expect
that this approach is not required in our case of rather dense bead positioning (typical bead difference
of 4 $\mu m$ and cell size around 100 $\mu m$), for future TFM applications it seems very interesting to explore in which situation
which combination of norms for the residue and the regularization term are most favorable.

We also emphasize the need for adaptations in the image processing procedures on wavy substrates. 
Because the point spread function for the marker beads can vary locally in this case, we cannot use
the standard cross correlation approach for bead tracking, but resort to a template matching procedure based on feature
vectors \cite{Legant2010}. We also take into account the anisotropy in the point spread function by
using a reduced weight for the z-component in the residue. Together with the use of the L1-norm for
the residue, this ensures a very good quality of our traction force reconstruction as validated by computer simulations.
Our analysis is facilitated by the use of topographic substrates with cylinder symmetry, which allows
us to use the slicing and meshing procedures shown in Fig.~\ref{femtfm_alpha_shapes_and_hexmesh}.

We next investigated the degree and orientation of cellular polarity for cardiac myofibroblasts,
which have been exposed to a wavy substrate geometry. We found that these cells strongly align and polarize perpendicular to the direction of maximal curvature, which has been reported before for other types of fibroblasts on rigid microcylinders.
We note that even round cells tend to align with the substrates features, presumabely because
the feature dimensions are such that every cell is affected by the topographical cues.
In addition to this observation we determined traction force maps for an exemplary set of cells and demonstrated the successful reconstruction of the 3D forces. We found that cells also balance their forces along the orientation of the pattern, even if at the same time forming a bridge between neighboring hills. Because our reconstruction is strongly regularized due to the noise issues with wavy substrates, the absolute values 
of the measured traction stresses are expected to be lower than the actual values,
as shown earlier with model-based traction force microscopy \cite{soine2015model}. In
fact, we expect the real traction stress to be closer to the kPa-range as used for the simulated data.

In the future, further advances in the experimental techniques as described here
(including faster acquisition of the image data, for example with a light sheet microscope)
should make a full statistical analysis feasible. It would be also interesting to conduct these experiments
with other cell types whose physiological function depends on curved or corrugated environment,
such as podocytes filtering the blood in kidney glomeruli.

\section{Acknowledgments}

This work was supported by MechanoSys (BMBF grant no. $0315501$) and the EcTop programme
of the cluster of excellence CellNetworks. BH, RM and USS would like to thank the Isaac Newton Institute for
Mathematical Sciences for its hospitality during the programme "Coupling geometric PDEs with physics for
cell morphology, motility and pattern formation" (CGPW02) supported by EPSRC Grant Number EP/K032208/1.
USS is a member of the Interdisciplinary Center for Scientific Computing (IWR) at Heidelberg. 

\section{Supplemental Information: FEM-implementation}

For a given traction pattern $\mathbf{t}(\mathbf{x})$, the direct boundary value problem (BVP) formulated in  Eq.~1
is now solved by means of the finite element method (FEM) as typically applied to elastic problems \cite{Braess_book,Ameen_book,Bonet1997}.
The first step in such a calculation is the transformation of the given equations into the weak form. Therefore we multiply the equation with an arbitrary field $\delta \mathbf{u}$ and integrate over the substrate volume $\Omega$:
\begin{equation}
	\int_{\Omega} (\nabla \delta \mathbf{u})^{T} : \mathrm{C} : \nabla \mathbf{u} ~dV = \int_{\partial \Omega} (\delta \mathbf{u})^T \mathbf{t} ~dS.
\end{equation} 
$\mathrm{C}$ represents the constant elasticity matrix for isotropic elastic materials, $\mathrm{C}_{ijkl} = \lambda \delta_{ij}\delta_{kl} + \mu (\delta_{ik}\delta_{jl} + \delta_{il}\delta_{jk})$. In the following we impose a discretization scheme, by partitioning the integral over $\Omega$ into smaller elements corresponding the generated substrate mesh. We further apply a 
local interpolation scheme. This has the objective to reduce the infinite dimensional space of displacement solutions to a finite dimensional subspace of nodal displacement values. The weak form for a single element reads:
\begin{equation}
\int_{\Omega^e} (\nabla \delta \mathbf{u}^e)^T : \mathrm{C} : \nabla \mathbf{u}^e ~dV = \int_{\partial \Omega^e} (\delta \mathbf{u}^e)^T \mathbf{t} ~dS.
\end{equation}
In our calculations, we consider hexahedral elements with eight nodal points at the element boundary. As an advantage of hexahedral elements, the volume and surface integration can be mapped to an integration over the unit cube parametrized by Cartesian coordinates, $(x_1,x_2,x_3)\rightarrow (\xi_1, \xi_2, \xi_3)$, while the coordinate transformation is determined by an individual Jacoby matrix calculated for each element. Thus the integration reads:
\begin{equation}
\int_{\tilde \Omega^e} (\nabla_{\xi} \delta \tilde{\mathbf{u}}^e)^T : \mathrm{C} : \nabla_{\xi} \tilde{\mathbf{u}}^e J_V ~d\tilde V =  \int_{\partial \tilde \Omega^e} \delta \tilde{\mathbf{u}}^e \mathbf{t} J_S ~d \tilde S. \label{element_virtual_work}
\end{equation}
$J$ represents the Jacobian and the tilde marks quantities with respect to the new coordinate $\boldsymbol \xi$. For subsequent interpolation of the displacement field we use interpolation functions $\Phi_n(\boldsymbol{\xi})$, which are based on an elementary set of linear shape functions. By interpolation of nodal values we can rewrite the displacement field, which now only dependents on a discrete number of degrees of freedom (DoFs) (in the hexahedral case used here, this is the number of nodes multiplied by the dimension, $8*3=32$):
\begin{equation}
\tilde{\mathbf{u}}^e (\boldsymbol{\xi})  = N(\boldsymbol{\xi}) \mathbf{u}^{e}_{node} =
\begin{pmatrix}
\Phi_1 & 0 & 0  & \dots & \Phi_n & 0 & 0 \\
0 & \Phi_1 & 0 &  \dots & 0 & \Phi_n & 0 \\
0 & 0 & \Phi_1 &  \dots & 0 & 0 & \Phi_n 
\end{pmatrix}
\begin{pmatrix}
u_1 \\
u_2 \\
u_3 \\
\vdots \\
u_n
\end{pmatrix}
\end{equation}
$\mathbf{u}^e_{node}$ is a vector of nodal displacement DoF values and $N(\boldsymbol{\xi})$ is the interpolation matrix, which interpolates the element displacement $\tilde{\mathbf{u}}^e(\boldsymbol{\xi})$ for a given configuration of nodal values $\mathbf{u}^e_{node}$. The same can be achieved for the virtual displacement $\delta\tilde{\mathbf{u}}^e(\boldsymbol{\xi})=N(\boldsymbol{\xi}) \delta \mathbf{u}^e_{node}$. Applying this to Eq.~\ref{element_virtual_work}, we obtain
\begin{equation}
	\delta \mathbf{u}^e_{node} \underbrace{\left[ \int_{\Omega} (\nabla_{\xi} N(\boldsymbol{\xi}))^T:\mathrm{C}:\nabla_\xi N(\boldsymbol{\xi}) J_V ~d\tilde V \right]}_{K^e} \mathbf{u}^e_{node} = \delta \mathbf{u}^e_{node} \underbrace{\left[\int_{\partial \Omega^e} N(\boldsymbol{\xi})^{T} \mathbf{t}   J_S ~dS \right]}_{\mathbf{R}^e}
\end{equation}
$K^e$ is called the element stiffness matrix and $\mathbf{R}^e$ the element load vector. Since $\delta \mathbf{u}^e_{node}$ is a vector of arbitrary values, we can reduce the problem to solving a linear algebraic system: 
\begin{equation}
	K^e \mathbf{u}^e_{node} = \mathbf{R}^e. \label{femtfm_FEM_local_system}
\end{equation}
The components of the element stiffness matrix and the load vector are calculated numerically. Here, the integrals have been solved by means of Gauss quadrature. In this way we calculate the stiffness matrix and load vector for each element. In a subsequent step, we assemble a global system that forms the domain $\Omega$:
\begin{equation}
	K \mathbf{U}_{node} = \mathbf{R}. \label{femtfm_FEM_global_system}
\end{equation}
$K$, $\mathbf{R}$, and $\mathbf{U}_{node}$ are the global stiffness matrix, the global load vector, and the global DoF vector. 
Nodal DoFs are only shared by neighboring elements and thus $K$ is a sparse matrix, while most of the matrix entries are zero. 
$K$ is further singular and hence not invertible, since we have still not introduced constraints to avoid rigid body motions.
Therefore, the system must be further restrained by incorporating appropriate boundary
conditions (BCs). For our direct BVP we consider the BCs illustrated in Fig.~2. The traction BC enters the system through the surface integral in the load vector $\mathbf{R}$. The zero stress BC leads to no constraints of the system. Only the remaining zero displacement BC applied to the bottom surfaces constrains Eq.~\ref{femtfm_FEM_global_system} and by forcing the displacement condition, we can reduce the system to
\begin{equation}
		K_{f} \mathbf{U_{f,node}} = \mathbf{R_f}  \label{femtfm_fem_reduced_system}
\end{equation}
where $K_{f}$ is the reduced non-singular stiffness matrix, $\mathbf{U}_{f,node}$ is the global vector of unconstrained (free) DoFs, and $\mathbf{R_f}$ is the corresponding load vector. We solve the system with respect to $\mathbf{U}_{f,node}$ by using the conjugated gradient (CG) method. Alternatively, it is also possible to directly invert $K$ by means of e.g. Gauss-elimination. In order to evaluate a displacement solutions at every position within the domain $\Omega$, we apply interpolation via the introduced functions $\Phi_n$ with respect to obtained global DoF configuration $\mathbf{U}$.

\newpage


\begin{thebibliography}{10}

\bibitem{Vogel2006}
Viola Vogel and Michael Sheetz.
\newblock {Local force and geometry sensing regulate cell functions.}
\newblock {\em Nature Reviews. Molecular Cell Biology}, 7(4):265--75, April
  2006.

\bibitem{Chen2008}
Christopher~S Chen.
\newblock {Mechanotransduction - a field pulling together?}
\newblock {\em Journal of Cell Science}, 121(Pt 20):3285--92, October 2008.

\bibitem{Humphrey2014}
Jay~D. Humphrey, Eric~R. Dufresne, and Martin~A. Schwartz.
\newblock Mechanotransduction and extracellular matrix homeostasis.
\newblock {\em Nature Reviews Molecular Cell Biology}, 15(12):802--812,
  December 2014.

\bibitem{Pelham1997}
Robert~J Pelham and Yu-Li Wang.
\newblock {Cell locomotion and focal adhesions are regulated by substrate
  flexibility.}
\newblock {\em Proceedings of the National Academy of Sciences of the United
  States of America}, 94:13661--5, June 1997.

\bibitem{Engler2004}
Adam Engler, Lucie Bacakova, Cynthia Newman, Alina Hategan, Maureen Griffin,
  and Dennis Discher.
\newblock Substrate {Compliance} versus {Ligand} {Density} in {Cell} on {Gel}
  {Responses}.
\newblock {\em Biophysical Journal}, 86(1):617--628, January 2004.

\bibitem{Nisenholz2014}
Noam Nisenholz, Kavitha Rajendran, Quynh Dang, Hao Chen, Ralf Kemkemer,
  Ramaswamy Krishnan, and Assaf Zemel.
\newblock Active mechanics and dynamics of cell spreading on elastic
  substrates.
\newblock {\em Soft Matter}, 10(37):7234--7246, August 2014.

\bibitem{McBeath2004}
Rowena McBeath, Dana~M Pirone, Celeste~M Nelson, Kiran Bhadriraju, and
  Christopher~S Chen.
\newblock Cell {Shape}, {Cytoskeletal} {Tension}, and {RhoA} {Regulate} {Stem}
  {Cell} {Lineage} {Commitment}.
\newblock {\em Developmental Cell}, 6(4):483--495, April 2004.

\bibitem{Engler2006}
Adam~J Engler, Shamik Sen, H~Lee Sweeney, and Dennis~E Discher.
\newblock {Matrix elasticity directs stem cell lineage specification.}
\newblock {\em Cell}, 126(4):677--89, August 2006.

\bibitem{Wen2014}
Jessica~H. Wen, Ludovic~G. Vincent, Alexander Fuhrmann, Yu~Suk Choi, Kolin~C.
  Hribar, Hermes Taylor-Weiner, Shaochen Chen, and Adam~J. Engler.
\newblock Interplay of matrix stiffness and protein tethering in stem cell
  differentiation.
\newblock {\em Nature Materials}, 13(10):979--987, October 2014.

\bibitem{schwarz_physics_2013}
Ulrich~S. Schwarz and Samuel~A. Safran.
\newblock Physics of adherent cells.
\newblock {\em Reviews of Modern Physics}, 85(3):1327--1381, August 2013.

\bibitem{Bischofs2005}
U.~S. Schwarz and I.~B. Bischofs.
\newblock Physical determinants of cell organization in soft media.
\newblock {\em Med. Eng. Phys.}, 27:763--72, 2005.

\bibitem{Kollmannsberger2011}
P.~Kollmannsberger, C.~M. Bidan, J.~W.~C. Dunlop, and P.~Fratzl.
\newblock {The physics of tissue patterning and extracellular matrix
  organisation: how cells join forces}.
\newblock {\em Soft Matter}, pages 9549--9560, 2011.

\bibitem{kraning2012controlling}
Casey~M Kraning-Rush and Cynthia~A Reinhart-King.
\newblock Controlling matrix stiffness and topography for the study of tumor
  cell migration.
\newblock {\em Cell adhesion \& migration}, 6(3):274--279, 2012.

\bibitem{Chen1997a}
C.~S. Chen.
\newblock {Geometric Control of Cell Life and Death}.
\newblock {\em Science}, 276(5317):1425--1428, May 1997.

\bibitem{Kilian2010}
Kristopher~a Kilian, Branimir Bugarija, Bruce~T Lahn, and Milan Mrksich.
\newblock {Geometric cues for directing the differentiation of mesenchymal stem
  cells.}
\newblock {\em Proceedings of the National Academy of Sciences of the United
  States of America}, 107(11):4872--7, March 2010.

\bibitem{Curtis1997}
Adam Curtis and Chris Wilkinson.
\newblock {Topographical control of cells.}
\newblock {\em Biomaterials}, 18(24):1573--83, December 1997.

\bibitem{Bettinger2009}
Christopher~J Bettinger, Robert Langer, and Jeffrey~T Borenstein.
\newblock {Engineering substrate topography at the micro- and nanoscale to
  control cell function.}
\newblock {\em Angewandte Chemie (International ed. in English)},
  48(30):5406--15, January 2009.

\bibitem{Kim2012}
Deok-Ho Kim, Paolo~P Provenzano, Chris~L Smith, and Andre Levchenko.
\newblock {Matrix nanotopography as a regulator of cell function}.
\newblock {\em The Journal of Cell Biology}, 197(3):351--360, April 2012.

\bibitem{dalby2007control}
Matthew~J Dalby, Nikolaj Gadegaard, Rahul Tare, Abhay Andar, Mathis~O Riehle,
  Pawel Herzyk, Chris~DW Wilkinson, and Richard~OC Oreffo.
\newblock The control of human mesenchymal cell differentiation using nanoscale
  symmetry and disorder.
\newblock {\em Nature materials}, 6(12):997--1003, 2007.

\bibitem{yim2007synthetic}
Evelyn~KF Yim, Stella~W Pang, and Kam~W Leong.
\newblock Synthetic nanostructures inducing differentiation of human
  mesenchymal stem cells into neuronal lineage.
\newblock {\em Experimental cell research}, 313(9):1820--1829, 2007.

\bibitem{dalby2003nucleus}
Matthew~J Dalby, Mathis~O Riehle, Stephen~J Yarwood, Chris~DW Wilkinson, and
  Adam~SG Curtis.
\newblock Nucleus alignment and cell signaling in fibroblasts: response to a
  micro-grooved topography.
\newblock {\em Experimental cell research}, 284(2):272--280, 2003.

\bibitem{Biela2009}
Sarah~a Biela, Yi~Su, Joachim~P Spatz, and Ralf Kemkemer.
\newblock {Different sensitivity of human endothelial cells, smooth muscle
  cells and fibroblasts to topography in the nano-micro range.}
\newblock {\em Acta Biomaterialia}, 5(7):2460--6, September 2009.

\bibitem{Ulbrich2011}
Stefan Ulbrich, Jens Friedrichs, Monika Valtink, Simo Murovski, Clemens~M
  Franz, Daniel~J M\"{u}ller, Richard H~W Funk, and Katrin Engelmann.
\newblock {Retinal pigment epithelium cell alignment on nanostructured collagen
  matrices.}
\newblock {\em Cells, tissues, organs}, 194(6):443--56, January 2011.

\bibitem{Corey2003}
Jospeh~M Corey and Eva~L Feldman.
\newblock {Substrate patterning: an emerging technology for the study of
  neuronal behavior}.
\newblock {\em Experimental Neurology}, 184:89--96, November 2003.

\bibitem{Svitkina1995}
T.M. Svitkina, Y.A. Rovensky, A.D. Bershadsky, and J.M. Vasiliev.
\newblock {Transverse pattern of microfilament bundles induced in
  epitheliocytes by cylindrical substrata}.
\newblock {\em Journal of Cell Science}, 108(2):735, February 1995.

\bibitem{Levina1996}
Elina~M Levina, Lidia~V Domnina, Yuri~A Rovensky, and Jury~M Vasiliev.
\newblock {Cylindrical Substratum Induces Different Patterns of Actin
  Microfilament Bundles in Nontransformed and in ras-Transformed
  Epitheliocytes}.
\newblock {\em Experimental Cell Research}, 229:159--165, 1996.

\bibitem{Biton2009}
Y~Y Biton and S~A Safran.
\newblock {The cellular response to curvature-induced stress}.
\newblock {\em Physical Biology}, 6:046010, 2009.

\bibitem{Sanz-Herrera2009}
Jos\'{e}~a Sanz-Herrera, Pedro Moreo, Jos\'{e}~M Garc\'{\i}a-Aznar, and Manuel
  Doblar\'{e}.
\newblock {On the effect of substrate curvature on cell mechanics.}
\newblock {\em Biomaterials}, 30(34):6674--86, December 2009.

\bibitem{charest2012fabrication}
Jonathan~M Charest, Joseph~P Califano, Shawn~P Carey, and Cynthia~A
  Reinhart-King.
\newblock Fabrication of substrates with defined mechanical properties and
  topographical features for the study of cell migration.
\newblock {\em Macromolecular bioscience}, 12(1):12--20, 2012.

\bibitem{Endlich2006}
Nicole Endlich and Karlhans Endlich.
\newblock {Stretch, tension and adhesion – Adaptive mechanisms of the actin
  cytoskeleton in podocytes}.
\newblock {\em European Journal of Cell Biology}, 85:229--234, 2006.

\bibitem{Cesa2007}
Claudia~M Cesa, N~Kirchgessner, D~Mayer, U~Schwarz, B~Hoffmann, and R~Merkel.
\newblock {Micropatterned silicone elastomer substrates for high resolution
  analysis of cellular force patterns.}
\newblock {\em The Review of Scientific Instruments}, 78:034301, March 2007.

\bibitem{Style2014}
Robert~W. Style, Rostislav Boltyanskiy, Guy~K. German, Callen Hyland,
  Christopher~W. MacMinn, Aaron~F. Mertz, Larry~A. Wilen, Ye~Xu, and Eric~R.
  Dufresne.
\newblock Traction force microscopy in physics and biology.
\newblock {\em Soft Matter}, 10(23):4047--4055, May 2014.

\bibitem{Plotnikov2014}
Sergey~V Plotnikov, Benedikt Sabass, Ulrich~S Schwarz, and Clare~M Waterman.
\newblock {\em {High-resolution traction force microscopy.}}, volume 123.
\newblock Elsevier Inc., 1 edition, January 2014.

\bibitem{Schwarz2015}
Ulrich~S Schwarz and J\'{e}r\^{o}me R~D Soin\'{e}.
\newblock {Traction force microscopy on soft elastic substrates: A guide to
  recent computational advances.}
\newblock {\em Biochimica et Biophysica Acta}, 1853(11):3095--3104, 2015.

\bibitem{Dembo1999}
M~Dembo and Yu-li Wang.
\newblock {Stresses at the Cell-to-Substrate Interface during Locomotion of
  Fibroblasts}.
\newblock {\em Biophysical Journal}, 76(4):2307--2316, 1999.

\bibitem{Merkel2007}
Rudolf Merkel, Norbert Kirchgessner, Claudia~M Cesa, and Bernd Hoffmann.
\newblock {Cell force microscopy on elastic layers of finite thickness.}
\newblock {\em Biophysical Journal}, 93(9):3314--23, November 2007.

\bibitem{DelAlamo2007}
Juan~C {Del Alamo}, Ruedi Meili, Baldomero Alonso-Latorre, Javier
  Rodr\'{\i}guez-Rodr\'{\i}guez, Alberto Aliseda, Richard~a Firtel, and Juan~C
  Lasheras.
\newblock {Spatio-temporal analysis of eukaryotic cell motility by improved
  force cytometry.}
\newblock {\em Proceedings of the National Academy of Sciences of the United
  States of America}, 104(33):13343--8, August 2007.

\bibitem{Butler2002}
James~P Butler, Iva~Marija Toli\'{c}-N\o~rrelykke, Ben Fabry, and Jeffrey~J
  Fredberg.
\newblock {Traction fields, moments, and strain energy that cells exert on
  their surroundings.}
\newblock {\em American Journal of Physiology. Cell Physiology}, 282:C595--605,
  March 2002.

\bibitem{Delanoe-Ayari2010}
H.~Delano\"{e}-Ayari, J.~Rieu, and M.~Sano.
\newblock {4D Traction Force Microscopy Reveals Asymmetric Cortical Forces in
  Migrating Dictyostelium Cells}.
\newblock {\em Physical Review Letters}, 105(24):2--5, December 2010.

\bibitem{Hur2012}
Sung~Sik Hur, Juan~C {Del \'{A}lamo}, Joon~Seok Park, Yi-Shuan Li, Hong~a
  Nguyen, Dayu Teng, Kuei-Chun Wang, Leona Flores, Baldomero Alonso-Latorre,
  Juan~C Lasheras, and Shu Chien.
\newblock {Roles of cell confluency and fluid shear in 3-dimensional
  intracellular forces in endothelial cells.}
\newblock {\em Proceedings of the National Academy of Sciences of the United
  States of America}, 109(28):11110--5, July 2012.

\bibitem{Legant2013}
Wesley~R Legant, Colin~K Choi, Jordan~S Miller, Lin Shao, Liang Gao, Eric
  Betzig, and Christopher~S Chen.
\newblock {Multidimensional traction force microscopy reveals out-of-plane
  rotational moments about focal adhesions.}
\newblock {\em Proceedings of the National Academy of Sciences of the United
  States of America}, 110(3):881--6, January 2013.

\bibitem{Yang2006}
Zhaochun Yang, Jeen-Shang Lin, Jianxin Chen, and James H-C Wang.
\newblock {Determining substrate displacement and cell traction fields-a new
  approach.}
\newblock {\em Journal of theoretical biology}, 242(3):607--16, October 2006.

\bibitem{Legant2010}
Wesley~R Legant, Jordan~S Miller, Brandon~L Blakely, Daniel~M Cohen, Guy~M
  Genin, and Christopher~S Chen.
\newblock {Measurement of mechanical tractions exerted by cells in
  three-dimensional matrices.}
\newblock {\em Nature Methods}, 7(12), November 2010.

\bibitem{steinwachs2015}
Julian Steinwachs, Claus Metzner, Kai Skodzek, Nadine Lang, Ingo Thievessen,
  Christoph Mark, Stefan M{\"u}nster, Katerina~E Aifantis, and Ben Fabry.
\newblock Three-dimensional force microscopy of cells in biopolymer networks.
\newblock {\em Nature methods}, 2015.

\bibitem{Maskarinec2009}
Stacey~A Maskarinec, Christian Franck, David~A Tirrell, and Guruswami
  Ravichandran.
\newblock {Quantifying cellular traction forces in three dimensions.}
\newblock {\em Proceedings of the National Academy of Sciences of the United
  States of America}, 106(52):22108--13, December 2009.

\bibitem{Toyjanova2014}
Jennet Toyjanova, Eyal Bar-Kochba, Cristina L\'{o}pez-Fagundo, Jonathan
  Reichner, Diane Hoffman-Kim, and Christian Franck.
\newblock {High Resolution, Large Deformation 3D Traction Force Microscopy.}
\newblock {\em PloS one}, 9(4):e90976, January 2014.

\bibitem{ulbricht2013cellular}
Anna Ulbricht, Felix~J Eppler, Victor~E Tapia, Peter~FM van~der Ven, Nico
  Hampe, Nils Hersch, Padmanabhan Vakeel, Daniela Stadel, Albert Haas, Paul
  Saftig, et~al.
\newblock Cellular mechanotransduction relies on tension-induced and
  chaperone-assisted autophagy.
\newblock {\em Current Biology}, 23(5):430--435, 2013.

\bibitem{hersch2013constant}
Nils Hersch, Benjamin Wolters, Georg Dreissen, Ronald Springer, Norbert
  Kirchge{\ss}ner, Rudolf Merkel, and Bernd Hoffmann.
\newblock The constant beat: cardiomyocytes adapt their forces by equal
  contraction upon environmental stiffening.
\newblock {\em Biology open}, 2(3):351--361, 2013.

\bibitem{Hampe2014}
Nico Hampe, Thorsten Jonas, Benjamin Wolters, Nils Hersch, Bernd Hoffmann, and
  Rudolf Merkel.
\newblock {Defined 2-D microtissues on soft elastomeric silicone rubber using
  lift-off epoxy-membranes for biomechanical analyses.}
\newblock {\em Soft matter}, 10(14):2431--43, 2014.

\bibitem{Gordan2008}
Ovidiu~D Gordan, Bo~N~J Persson, Claudia~M Cesa, Dirk Mayer, Bernd Hoffmann,
  Sabine Dieluweit, and Rudolf Merkel.
\newblock {On pattern transfer in replica molding.}
\newblock {\em Langmuir : the ACS Journal of Surfaces and Colloids},
  24(13):6636--9, June 2008.

\bibitem{Edelsbrunner1983}
Herbert Edelsbrunner, David~G. Kirkpatrick, and Raimund Seidel.
\newblock {On the Shape of a Set of Points in the Plane}.
\newblock {\em IEEE Transactions on Information Theory}, IT-29(4):551--559,
  1983.

\bibitem{cgal}
{CGAL, Computational Geometry Algorithms Library}.

\bibitem{gmsh}
{Gmsh: a three-dimensional finite element mesh generator with built-in pre- and
  post-processing facilities}.

\bibitem{Braess_book}
Dietrich Braess.
\newblock {\em Finite Elemente Theorie: schnelle L\"oser und Anwendungen in der
  Elastizitätstheorie}.
\newblock Springer Verlag, Berlin, Germany, 4th ed. edition, 2007.

\bibitem{Ameen_book}
Mohammed Ameen.
\newblock {\em Computational Elasticity}.
\newblock Alpha Science Int. Ltd., Harrow, U.K., 2005.

\bibitem{Bonet1997}
Javier Bonet and Richard~D Wood.
\newblock {\em {Nonlinear Continuum Mechanics for Finite Element Analysis}}.
\newblock Cambridge University Press, 1997.

\bibitem{Bangerth2007}
Wolfgang Bangerth, Ralf Hartmann, and Guido Kanschat.
\newblock {deal.II---A general-purpose object-oriented finite element library}.
\newblock {\em ACM Transactions on Mathematical Software}, 33(4), August 2007.

\bibitem{Tikhonov1977}
Andrey~N Tikhonov and Vasiliy~Y Arsenin.
\newblock {\em {Solution of Ill-Posed Problems}}.
\newblock John Wiley \& Sons, New York, 1977.

\bibitem{Vogel2002}
Curtis~R Vogel.
\newblock {\em {Computational Methods for Inverse Problems}}.
\newblock Society for Industrial and Applied Mathematics, Philadelphia, 2002.

\bibitem{Schwarz2002}
Ulrich~S Schwarz, Nathalie~Q Balaban, Daniel Riveline, Alexander~D Bershadsky,
  Benjamin Geiger, and Sam Safran.
\newblock {Calculation of Forces at Focal Adhesions from Elastic Substrate
  Data: The Effect of Localized Force and the Need for Regularization}.
\newblock {\em Biophysical Journal}, 83:1380--1394, September 2002.

\bibitem{Sabass2008}
Benedikt Sabass, Margaret~L Gardel, Clare~M Waterman, and Ulrich~S Schwarz.
\newblock {High resolution traction force microscopy based on experimental and
  computational advances.}
\newblock {\em Biophysical Journal}, 94:207--220, January 2008.

\bibitem{Maronna2006}
Ricardo~A Maronna, R~Douglas Martin, and V$\backslash$'ictor~J Yohai.
\newblock {\em {Robust Statistics - Theory and Methods}}.
\newblock John Wiley \& Sons, Ltd, 2006.

\bibitem{NumericalRecipes2007}
William H~[Hrsg.] Press, Saul A~[Hrsg.] Teukolsky, William T~[Hrsg.]
  Vetterling, and Brian P~[Hrsg.] Flannery, editors.
\newblock {\em {Numerical recipes}}.
\newblock Cambridge Univ. Press, Cambridge [u.a.], 3rd ed. edition, 2007.

\bibitem{Lukyanenko2012}
D~V Lukyanenko and A~G Yagola.
\newblock {Using Parallel Computing for Solving Multidimensional Ill-posed
  Problems}.
\newblock In Y~Wang, A~G Yagola, and Y~Changchun, editors, {\em Computational
  Methods for Applied Inverse Problems}, pages 49--81. De Gruyter, Boston, MA,
  USA, 2012.

\bibitem{Balaban2001}
Nathalie~Q Balaban, Ulrich~S Schwarz, Daniel Riveline, Polina Goichberg, Gila
  Tzur, Ilana Sabanay, Diana Mahalu, Sam Safran, Alexander Bershadsky, Lia
  Addadi, and Benjamin Geiger.
\newblock {Force and focal adhesion assembly: a close relationship studied
  using elastic micropatterned substrates.}
\newblock {\em Nature Cell Biology}, 3(5):466--72, May 2001.

\bibitem{Fu2010}
Jianping Fu, Yang-Kao Wang, Michael~T Yang, Ravi~a Desai, Xiang Yu, Zhijun Liu,
  and Christopher~S Chen.
\newblock {Mechanical regulation of cell function with geometrically modulated
  elastomeric substrates}.
\newblock {\em Nature Methods}, 7(9):733--736, August 2010.

\bibitem{Hur2009}
Sung~Sik Hur, Yihua Zhao, Yi-Shuan Li, Elliot Botvinick, and Shu Chien.
\newblock {Live Cells Exert 3-Dimensional Traction Forces on Their Substrata.}
\newblock {\em Cellular and Molecular Bioengineering}, 2(3):425--436, September
  2009.

\bibitem{kristal2013metastatic}
R~Kristal-Muscal, Liron Dvir, and Daphne Weihs.
\newblock Metastatic cancer cells tenaciously indent impenetrable, soft
  substrates.
\newblock {\em New Journal of Physics}, 15(3):035022, 2013.

\bibitem{DvirWeihs2015}
Liron Dvir, Ronen Nissim, Martha~B Alvarez-Elizondo, and Daphne Weihs.
\newblock Quantitative measures to reveal coordinated cytoskeleton-nucleus
  reorganization during in vitro invasion of cancer cells.
\newblock {\em New Journal of Physics}, 17(4):043010, 2015.

\bibitem{Shtengel2009}
Gleb Shtengel.
\newblock {Interferometric fluorescent super-resolution microscopy resolves 3D
  cellular ultrastructure}.
\newblock {\em Proceedings of the National Academy of Sciences of the United
  States of America}, 106(9):3125--3130, 2009.

\bibitem{Xu2012}
Ke~Xu, Hazen~P Babcock, and Xiaowei Zhuang.
\newblock {Dual-objective STORM reveals three-dimensional filament organization
  in the actin cytoskeleton.}
\newblock {\em Nature Methods}, 9(2):185--8, February 2012.

\bibitem{Aquino2011}
Daniel Aquino, Andreas Sch\"{o}nle, Claudia Geisler, Claas Middendorff,
  Christian~A Wurm, Yosuke Okamura, Thorsten Lang, Stefan~W Hell, and Alexander
  Egner.
\newblock {Two-color nanoscopy of three-dimensional volumes by 4Pi detection of
  stochastically switched fluorophores}.
\newblock {\em Nature Methods}, 8(4):353--359, 2011.

\bibitem{Hansen1992}
Per~Christian Hansen.
\newblock {ANALYSIS OF DISCRETE ILL-POSED PROBLEMS BY MEANS OF THE L-CURV}.
\newblock {\em SIAM}, 34(4):561--580, 1992.

\bibitem{han2015traction}
Sangyoon~J Han, Youbean Oak, Alex Groisman, and Gaudenz Danuser.
\newblock Traction microscopy to identify force modulation in subresolution
  adhesions.
\newblock {\em Nature methods}, 12(7):653--656, 2015.

\bibitem{soine2015model}
J{\'e}r{\^o}me~RD Soin{\'e}, Christoph~A Brand, Jonathan Stricker, Patrick~W
  Oakes, Margaret~L Gardel, and Ulrich~S Schwarz.
\newblock Model-based traction force microscopy reveals differential tension in
  cellular actin bundles.
\newblock {\em PLoS Comput Biol}, 11(3):e1004076, 2015.

\end{thebibliography}

\end{document}